\documentclass{emulateapj} 
\usepackage{amsmath} 
\usepackage{apjfonts} 
\usepackage{amssymb} 
\usepackage{xspace}
\usepackage{hyperref} 
\usepackage{natbib} 
\usepackage{multirow}
\usepackage{graphicx}
\usepackage{epstopdf}
\usepackage{epsf}
\usepackage{subfigure}
\usepackage[usenames,dvipsnames]{xcolor}

\def\gsim{\gtrsim}
\def\lsim{\lesssim}    

\begin{document} 
\shorttitle{} 
\shortauthors{Avestruz et al.}
\accepted{The Astrophysical Journal} 
\slugcomment{The Astrophysical Journal, accepted}

\title{Testing X-ray Measurements of Galaxy Cluster Outskirts \\ with
  Cosmological Simulations} 

\author{ Camille Avestruz\altaffilmark{1,2} }
\author{ Erwin T. Lau\altaffilmark{1,2} }
\author{ Daisuke Nagai\altaffilmark{1,2,3} }
\author{ Alexey Vikhlinin\altaffilmark{4}}
\affil{
$^1${Department of Physics, Yale University, New Haven, CT
  06520, U.S.A.; \href{mailto:camille.avestruz@yale.edu}{camille.avestruz@yale.edu}} \\
$^2${Yale Center for Astronomy \& Astrophysics, Yale
  University, New Haven, CT 06520, U.S.A.} \\
$^3${Department of Astronomy, Yale University, New Haven,
  CT 06520, U.S.A.}  \\
$^4${Harvard-Smithsonian Center for Astrophysics, 60
  Garden Street, Cambridge, MA 02138, U.S.A.}
}

  
\begin{abstract} 
  The study of galaxy cluster outskirts has emerged as one of the new
  frontiers in extragalactic astrophysics and cosmology with the
  advent of new observations in X-ray and microwave.  However, the
  thermodynamic properties and chemical enrichment of this diffuse and
  azimuthally asymmetric component of the intra-cluster medium are
  still not well understood. This work, for the first time,
  systematically explores potential observational biases in these
  regions. To assess X-ray measurements of galaxy cluster properties
  at large radii ($>{R}_{500c}$), we use mock {\em Chandra} analyses
  of cosmological galaxy cluster simulations. The pipeline is
  identical to that used for {\em Chandra} observations, but the
  biases discussed in this paper are relevant for all X-ray
  observations outside of ${R}_{500c}$.  We find the following from
  our analysis: (1) filament regions can contribute as much as 
    $50\%$ at ${R}_{200c}$ to the emission measure, (2) X-ray
  temperatures and metal abundances from model fitted mock X-ray
  spectra in a multi-temperature ICM respectively vary to the
  level of $10\%$ and $50\%$, (3) resulting density profiles vary to
  within $10\%$ out to ${R}_{200c}$, and gas mass, total mass,
  and baryon fractions vary all to within a few percent, (4) the bias
  from a metal abundance extrapolated a factor of 5 higher than the
  true metal abundance results in total mass measurements biased high
  by $20\%$ and total gas measurements biased low by $10\%$ and (5)
  differences in projection and dynamical state of a cluster can lead
  to gas density slope measurements that differ by a factor of $15\%$
  and $30\%$, respectively.  The presented results can partially
  account for some of the recent gas profile measurements in cluster
  outskirts by e.g., {\em Suzaku}.  Our findings are pertinent to
  future X-ray cosmological constraints with cluster outskirts, which
  are least affected by non-gravitational gas physics, as well as to
  measurements probing gas properties in filamentary structures.
\end{abstract}


\section{Introduction}


Clusters of galaxies are the largest gravitationally bound objects in
our universe.  These objects are massive enough to probe the tension
between dark matter and dark energy in structure formation, making
them powerful cosmological tools. Cluster-based cosmological tests use
observed cluster number counts, the precision of which depends on how
well observable-mass relations can constrain cluster masses.  Finely
tuned observable-mass relations require an understanding of the
thermal and chemical structure of galaxy clusters out to large radii,
as well as an understanding of systematic biases that may enter into
observational measurements of the intracluster medium (ICM).

A new area of study lies in studying the ICM in the outskirts of
clusters, a region where the understanding of ICM physics remains
incomplete \citep[e.g., see][for an overview]{reiprich_etal13}.
Cluster outskirts can be considered to be a cosmic melting pot, where
infalling substructure is undergoing virialization and the associated
infalling gas clumps are being stripped due to ram pressure,
depositing metal-rich gas in the cluster atmosphere, changing thermal
and chemical structure of the ICM.  Measurements at these regions will
allow us to form a more complete assessment of the astrophysical
processes, e.g. gas stripping and quenching of star formation in
infalling galaxies, that govern the formation and evolution of galaxy
clusters and their galaxies.

Gas properties in cluster outskirts are also more suitable for
cosmological measurements; the effects of dissipative
non-gravitational gas physics, such as radiative cooling and feedback,
are minimal in the outskirts compared with the inner regions. Gas
density measurements in cluster outskirts will also help constrain the
baryon budget.  While recent X-ray cluster surveys have provided an
independent confirmation of cosmic acceleration and tightened
constraints on the nature of dark energy
\citep{allen_etal08,vikhlinin_etal09}, the best X-ray data, circa
2009, were limited to radii within half of the virial radius, and
cluster outskirts were largely unobserved.

Recent measurements from the {\em Suzaku} X-ray satellite have
pioneered the study of the X-ray emitting ICM in cluster outskirts
beyond ${R}_{200c}$\lastpagefootnotes\footnote{${R}_{200c}$ is the radius of the cluster
  enclosing an average matter density 200 times the critical density
  of the Universe.}  \citep[e.g.,][]{bautz_etal09,reiprich_etal09,
  hoshino_etal10,kawaharada_etal10}.  These measurements had
unexpected results in both entropy and enclosed gas mass
fraction at large radii.  Entropy profiles from {\em Suzaku} data were
significantly flatter than theoretical predictions from hydrodynamical
simulations \citep{george_etal09,walker_etal12}, and the enclosed gas
mass fraction from gas mass measurements of the Perseus cluster
exceeded the cosmic baryon fraction \citep{simionescu_etal11}.  These
results suggest that gasesous inhomogeneities might cause an
overestimate in gas density at large radii. {\em Suzaku} observations
and subsequent studies demonstrated that measurements of the ICM in
cluster outskirts are complicated by contributions from the cosmic
X-ray background, excess emission from gasesous substructures 
\citep{nagaiandlau_11,zhuravleva_etal13,vazza_etal13,roncarelli_etal13},
and poorly understood ICM physics in the outskirts.  However, analyses
of cluster outskirts with both {\em Planck} Sunyaev-Zel'dovich
and {\em ROSAT} X-ray measurements found otherwise
\citep{eckert_etal13a,eckert_etal13b}.  The mysteries surrounding
these X-ray measurements of cluster outskirts are still unsettled due
to a number of observational systematic uncertainties, such as X-ray
background subtraction.

The latest deep {\em Chandra} observations of the outskirts of galaxy
cluster Abell 133, a visually relaxed cluster with X-ray temperature
$T_X=4.1$~keV, have given us an unprecedented opportunity to probe the
ICM structure in cluster outskirts (Vikhlinin~et~al., in prep.).  First,
the high sub-arcsec angular resolution allows for efficient point
source removal, the isolation of the cosmic X-ray background, and the
detection of small-scale clumps, which dominate the X-ray flux at
large radii.  Second, the long exposure time of 2.4~Msec also enables
direct observations of the filamentary distribution of gas out to
${R}_{200c}$.  

To address potential biases that could affect observed ICM properties
at large radii, we use mock {\em Chandra} observations from
high-resolution cosmological simulations of galaxy clusters.  The goal
of this paper is to characterize the properties of diffuse components
of the X-ray emitting ICM in the outskirts of galaxy clusters, and to
assess the implications for X-ray measurements.  Our analysis focuses
on the effects that come from a combination of limited spectral
resolution and low photon counts in cluster outskirts, where spectral
contributions from metal line emissions become more significant than
those from thermal bremsstrahlung emissions.

Our approach is similar in spirit as in \citet{rasia_etal08} which
pioneered the study of systematics on abundance measurements on the
ICM by analyzing the spectral properties of mock plasma spectra
generated with XSPEC.  We extend the study of XSPEC generated mock
spectra to the outskirt regions.  Our results are valid for all
observations of spectra from plasma with a multicomponent temperature.

Specifically, we look at the accuracy of metal abundance measurements
from spectral fitting and their consequential effects on the
measurements of projected X-ray temperature, emission measure,
deprojected temperature, gas density, gas mass, and hydrostatic mass
derived from these profiles.  We explore potential biases by (1)
measuring contributions of clumps and filaments to emission
measurements, (2) testing the modeling of bremsstrahlung and metal
line emissions in low density regions, and (3) checking for line of
sight dependence in the density slope in cluster outskirts and
differences in observed ICM profiles for an unrelaxed cluster.

This work will be useful in assessing the robustness of X-ray
measurements beyond ${R}_{500c}$, as we test both the spectral
fitting of X-ray photons in low density regions, and the validity of
fitting formulae that are used to reconstruct ICM profiles.

Our paper is organized as follows: in Section~\ref{sec:methods} we
describe the simulations we used and the mock {\em Chandra} analysis
pipeline.  We present our results in Section~\ref{sec:results}, and
give our summary and discussion in Section~\ref{sec:summary}.


\section{Methodology}
\label{sec:methods}

\subsection{Cosmological simulations}

We use a sample of high-resolution cosmological simulations of
cluster-sized systems from \citet[][hereafter N07]{nagai_etal07a,
  nagai_etal07b} that assumes a flat $\Lambda$CDM universe with
$\Omega_m=1-\Omega_\Lambda=0.3$, $\Omega_b=0.04286$, $h=0.7$, and
$\sigma_8=0.9$.  The Hubble constant is defined as $100h$ km~s$^{-1}$
Mpc$^{-1}$, and $\sigma_8$ is the mass variance within spheres of
radius $8\,h^{-1}$Mpc.  We simulate these clusters with the Adaptive
Refinement Tree (ART) $N$-body+gas-dynamics code
\citep{kravtsov_99,kravtsov_etal02,rudd_etal08}, an Eulerian code with
spatial and temporal adaptive refinement, and non-adaptive mass
refinement \citep{klypin_etal01}.  

Our simulations include radiative cooling, star formation, metal
enrichment, stellar feedback, and energy feedback from supermassive
black holes (see Avestruz~et~al., in~prep. for more details).
Briefly, we seed black hole particles with a seed mass of
$10^5h^{-1}M_{\odot}$ in dark matter halos with
$M_{500c}>2\times10^{11}h^{-1}M_{\odot}$, and allow
the black holes to accrete gas according to a modified Bondi accretion
model with a density dependent boost factor \citep{boothandschaye_09}.
Black holes feed back a fraction of the accreted rest mass energy in
the form of thermal energy,
\begin{equation}
  \Delta E_{fb}=\epsilon_r\epsilon_f\Delta M_{BH}
\end{equation}
where $\epsilon_r=0.1$ is the radiative efficiency for a Schwarzschild
black hole undergoing radiatively efficient Shakura \& Sunyaev
accretion \citep{shakuraandsunyaev_73}, and $\epsilon_f=0.2$ is the
feedback efficiency parameter tuned to regulate the $z=0$ black hole
mass.

We focus on two X-ray luminous clusters from the N07 sample to study
the effects of metallicity, line of sight, and dynamical state on
X-ray measurements in cluster outskirts. The core-excised X-ray
temperatures of CL6 and CL7 are $T_X=4.18$ and $T_X=4.11$,
respectively.  The cluster mass and radii are
  $M_{500c}=1.43\times10^{14}h^{-1}M_\odot$ and
    $R_{500c}=626.5\,h^{-1}\text{kpc}$ for CL6, and
    $M_{500c}=1.37\times10^{14}h^{-1}M_\odot$ and
    $R_{500c}=618.6\,h^{-1}\text{kpc}$ for CL7.  Each cluster is
  selected from simulation box with size
  ${L}_{\mathrm{box}}=80\,h^{-1}$~Mpc, run on a $128^3$ uniform grid
  with 8 levels of refinement, and has peak spatial resolution of
  $\approx2.4h^{-1}$~kpc.  The corresponding dark matter particle mass
  in our regions of interest is
$2.7\times10^8h^{-1}{M}_{\odot}$.  CL6 is an unrelaxed cluster that
experienced a 1:1 major merger at $z\approx0.6$, and CL7 is a typical
relaxed cluster which has not undergone a major merger for $z<1$.  
We refer the reader to N07 for further details.

\subsection{Mock {\em Chandra} Pipeline}

Using our simulated cluster sample, we create realistic mock X-ray
maps of simulated clusters by (1) generating an X-ray flux map, and
(2) convolving the flux map with the {\em Chandra} response files to
create photon maps.  Below we outline our mock {\em Chandra} pipeline,
where details can be found in \citet{nagai_etal07b}.

\subsubsection{X-ray Flux maps}

We generate X-ray flux maps for each of the simulated clusters along
three orthogonal line-of-sight projections.  To calculate the flux
map, we project X-ray emission from hydrodynamical cells along each
line of sight within $3\times {R}_{\mathrm{vir}}$ ($\approx
4\times{R}_{200c}$ for the clusters in this work) of the cluster
center.  The X-ray emission as a function of energy from a single hydrodynamical cell of
volume $\Delta V_i$ is given as,
\begin{equation}
j_{E,i}= n_{e,i}n_{p,i}\Lambda_{E}(T_i,Z_i,z)\Delta V_i,
\end{equation}
where $n_{e,i}$, $n_{p,i}$, $T_i$, and $Z_i$ are respectively the
electron number density, proton number density, temperature, and
metallicity in that cell volume.  We compute the X-ray emissivity,
$\Lambda_E(T_i, Z_i, z)$, using the MEKAL plasma code
\citep{mewe_etal85,kaastra_etal93,liedahl_etal95} and the
  solar abundance table from \citet{andersandgrevesse_89}.  We multiply the
plasma spectrum by the photoelectric absorption corresponding to a
hydrogen column density of $N_{\mathrm{H}}=2\times 10^{20}$~cm$^{-2}$.

We include results from four different flux maps: (1)
$Z_\mathrm{sim}$, (2) $Z_{0.5}$, (3) $Z_{0.05}$, and (4)
$Z_{0.5}T_{X,1\text{ keV}}$.  We generate the $Z_\mathrm{sim}$ flux
map from X-ray spectra using all hydrodynamic values directly output
from our simulation runs.  To isolate the effects of metallicity on
our simulated spectra, we use $Z_i=0.5 {Z}_{\odot}$ for all $i$-th
cell elements in our $Z_{0.5}$ flux maps and a constant
$Z_i=0.05\mathrm{Z}_{\odot}$ for all $i$-th cell elements in our
$Z_{0.05}$ flux maps. Current observations do not yet probe metal
abundance profiles in cluster outskirts, and often extrapolate the
abundance at radii beyond ${R}_{500c}$.  Given the typical range of
metallicity that observers have found near ${R}_{500c}$
($Z\approx0.3\mathrm{Z}_\odot$)
\citep[e.g.,][]{matsushita_etal07,leccardi_molendi08,komiyama_etal09,werner_etal13},
we choose these two values for $Z_i$ to bracket a potential range of
metallicities in the outskirts.  The order of magnitude difference
probes how X-ray measurements of the ICM and inferred quantities might
be affected by an incorrect extrapolation or mis-measurement of both a
metal rich and metal poor environment.

We remove the effects of a multi-temperature medium by fixing both the
metal abundance to $Z_i=0.5{Z}_\odot$, and the temperature to $T_X=1$
keV in the $Z_{0.5}T_{X,1\text{ keV}}$ map, which is a characteristic
temperature in the outskirts of a galaxy cluster in our probed mass
range.

\subsubsection{Mock {\em Chandra} Photon Maps and Spectra}
\label{sec:mockmaps_and_spectra}

To simulate mock {\em Chandra} data, we convolve the emission spectrum
with the response of the {\em Chandra} front-illuminated CCDs ACIS
(both the {\em Chandra} auxiliary response file (ARF) and
redistribution matrix file (RMF)) and draw a number of photons at each
position and spectral channel from the corresponding Poisson
distribution.  The photon maps for our mock {\em Chandra} data has an
exposure time of $2$~Msec, comparable to the $2.4$~Msec deep {\em
  Chandra} observations of Abell 133.  Since we are interested in
probing physical causes for biased X-ray measurements, we do not
include any simulated backgrounds.

From our mock photon maps, we generate images in the $0.7-2$~keV band,
and use them to identify and mask out clumps from our analysis.  We
use the wavelet decomposition algorithm described in
\citet{vikhlinin_etal98}.  This procedure is consistent with the
small-scale clump removal performed by observers.

\begin{figure*} 
  \centering
  \includegraphics[width=6in]{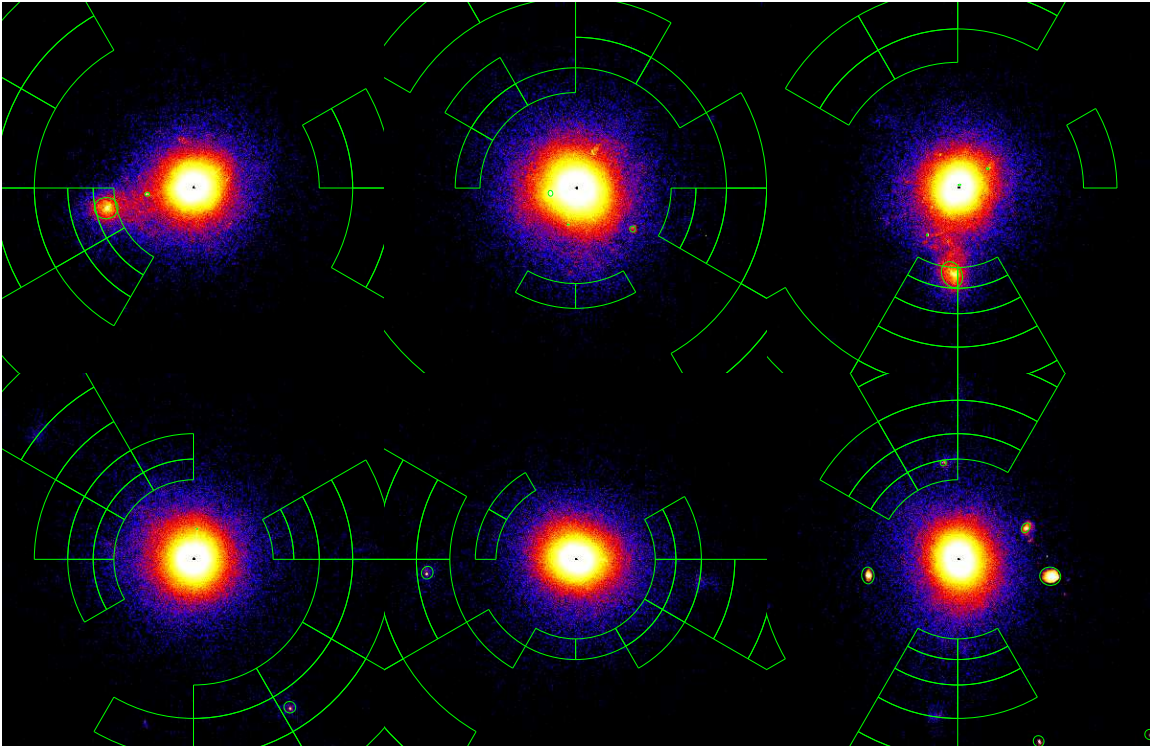}
  \caption{Top: Mock {\em Chandra} maps of CL6 in $0.7-2$~keV band
    viewed from left to right along the x, y, and z axes. Bottom: Mock
    {\em Chandra} maps of CL7 in $0.7-2$~keV band. Each region is
    5~Mpc across. Identified clumps and filamentary structures are
    respectively outlined by green ellipses annular sections. Both of
    these regions are excluded from our final analysis.}
  \label{fig:img_map}
\end{figure*}

\begin{figure} 
  \centering
  \includegraphics[width=3.5in]{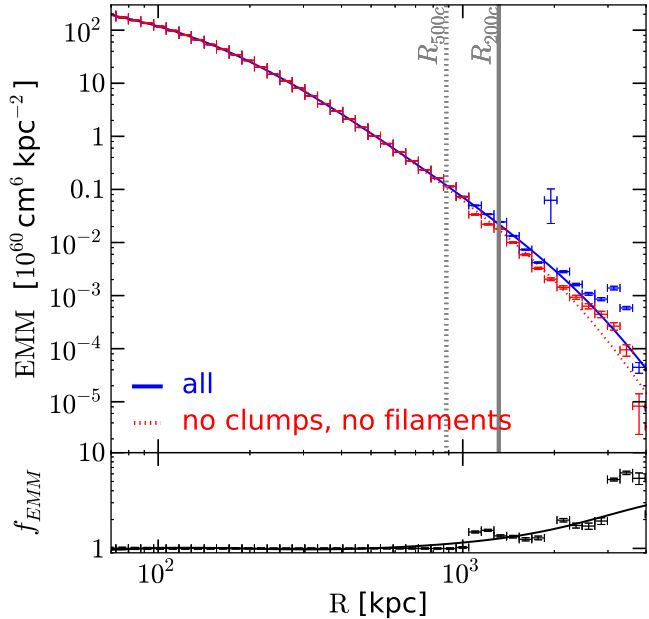}
  \caption{We show the effects of clump and filament removal on the
    emission measure profile for a sample projection for CL7 in the y
    direction.  Profiles with both clumps and filaments are shown in
    the solid lines, no clumps in the dash-dotted lines, and no clumps
    nor filaments in the dotted lines.  The ratio between solid
    (dash-dotted) and dotted is shown in the excess panel.
    Substructure in this cluster can boost the best fit emission
    measure by up to a factor of 3, and binned emission measure
      by up to a factor of 6 in the filamentary regions.}
  \label{fig:EMMxyz}
\end{figure}

Detected clumps are within a limited flux of
$\sim3\times10^{-15}$~erg~s$^{-1}$~cm$^{-2}$ in the $0.7-2$~keV energy
band, corresponding to a luminosity of $\sim
1.5\times10^{42}$~erg~s$^{-1}$ for clusters at $z_{\mathrm{obs}}=0.06$,
the observational redshift of our clusters.  Any remaining
  small-scale structures fall below our clump detection limit. The total
gas mass in excluded clumps is no more than a few percent of the total
enclosed gas mass.

X-ray emission in cluster outskirts also receive azimuthally
asymmetric contributions from filamentary structures. Similar to the
analysis of X-ray emission in A133 outskirts, after clump removal, we
identify filamentary regions as annular sections with an average flux
that exceeds the average flux within the entire annulus. We exclude
the filamentary contribution from our final analysis. We include a
sample image map with identified clumps and filaments in
Figure~\ref{fig:img_map}.

From the mock {\em Chandra} photon map, we extract a grouped spectrum
from concentric annular bins centered on the main X-ray peak.  Annular
regions are defined with $r_{\mathrm{out}}/r_{\mathrm{in}}=1.25$,
where $r_{\mathrm{out}}$ and $r_{\mathrm{in}}$ are the outer and inner
edges of the annular bin, respectively.  The spectra exclude photons
from regions corresponding to clumps and filaments.  For each
spectrum, we perform a single-temperature fit in the $0.1-10$~keV band
using {XSPEC} version 12.8 with the MEKAL model plasma code and a
C-statistic fit \citep{cash_79} to extract the projected X-ray
temperature ${T}_{X,\mathrm{proj}}(r)$ and abundance
${Z}_{\mathrm{proj}}(r)$.  We use the same {\em Chandra} ARF and RMF
files to fit a model spectrum to the mock photon spectra.
Temperature, metal abundance, and spectral normalization are the only
free parameters in the fit.

\subsubsection{Analysis of the ICM}

To compare with observed {\em Chandra} data, we reconstruct the emission
measure profile and repeat the deprojection analysis by
\citet{vikhlinin_etal06}.  This analysis reconstructs the spherically
averaged density and temperature profiles of the intracluster medium
with 3D analytic models.  We also estimate the total cluster mass
assuming hydrostatic equilibrium from the reconstructed 3D density and temperature profiles.

To calculate the projected emission measure profile ${EMM}=\int
n_en_pdl$, we use two main input ingredients: (1) the azimuthally
averaged X-ray surface brightness profile $XSB$, and (2) the projected
2D X-ray temperature $T_{X,\mathrm{proj}}$ and metallicity profiles $Z_{\mathrm{proj}}$ described in
Section~\ref{sec:mockmaps_and_spectra}.  
We can relate the deprojected $EMM$ to the two ingredients as follows,
\begin{equation}\label{eqn:cnt2emm}
XSB=\int dV\;n_en_p\Lambda(T,Z) 
\end{equation}
We show the emission measure profile for CL7 along the y-axis, where
filamentary features are most prominent in Figure~\ref{fig:EMMxyz},
and the corresponding effect of clump and filament removal from the
analysis.  The solid blue line corresponds to the emission measure
profile that include clumps and filamentary substructure, and dotted
line corresponds to the emission measure profile that excludes the
substructure.  The bottom panel shows the ratio between the profile
with substructure to the profile without, and illustrates the extent
of the excess emission in cluster outskirts.  For this cluster, the
excess of the best-fit profile can be as much as a factor of 2 at
${R}_{200c}$ to a factor of 3 at $3\times{R}_{200c}$.  The errorbars
correspond to the binned data, and the excess in individual bins are
as high as a factor of 6 in the filamentary regions.  Excess emission
measure from a substructure is visible at $R\approx2$ kpc.

We extract the X-ray surface brightness profiles from photons in the
$0.7-2$~keV band in concentric annuli of
$r_{\mathrm{out}}/r_{\mathrm{in}}=1.1$, giving us a count rate in each
annular bin.  We use the projected X-ray temperature and metallicity
to calculate the MEKAL model plasma emissivity,
$\Lambda(T_{X,\mathrm{proj}}(r_{\mathrm{proj}}),Z_{\mathrm{proj}}(r_{\mathrm{proj}}),z)$.
With the relation given by Equation~(\ref{eqn:cnt2emm}), we use interpolated
values of the model plasma emissivity to convert the count rate into a
projected emission measure.

The deprojection process is fully described in
\citet{vikhlinin_etal06}, and can be summarized as follows.  The
analytic form of our 3D gas density model is given by,
\begin{align}\label{eqn:npne}
n_pn_e = & \frac{n^2_0\left(
    r/r_c\right)^{-\alpha}}{\left(1+r^2/r^2_c\right)^{3\beta-\alpha/2}}
\frac{1}{\left(1+r^\gamma/r_s^\gamma\right)^{\epsilon/\gamma}}\nonumber \\
&+\frac{n^2_{02}}{\left(1+r^2/r^2_{c2}\right)^{3\beta_2}}
\end{align}
This is a modification of the $\beta$-model that allows for separate fitting of the
gas density slope at small, intermediate, and large radii.

The analytic form of our 3D temperature profile is a product of two
terms that describe key features of in observed projected temperature
profiles \citep{vikhlinin_etal05},
\begin{equation}
{T}_\mathrm{3D}(r) = t(r)\times t_{\mathrm{cool}}(r)
\end{equation}
The first term takes into account the temperature decline at large
radii with a broken power law with transition radius, $r_t$,
\begin{equation}
t(r) = {T}_0\frac{\left(r/r_t\right)^{-a}}{\left(1+(r/r_t)^b\right)^{c/b}}
\end{equation}
with $a$ and $b$ as free parameters.  The second term models the
temperature decline in the central region due to radiative cooling,
\begin{equation}
t_{\mathrm{cool}}=\frac{\left(r/r_{\mathrm{cool}}\right)^{a_{\mathrm{cool}}}
  +{T}_{\mathrm{min}}/{T}_0}{\left(r/r_{\mathrm{cool}}\right)^{a_{\mathrm{cool}}+1}},
\end{equation}
We project the 3D model of the temperature profile, weighting multiple
temperature components using the algorithm from
\citet{vikhlinin_etal06}.  We then fit the projected model to the
projected X-ray temperature from our mock data, fitting in the radial
range between $0.1\leq r/R_{200c} \leq 3$.

The 3D models for gas density and temperature profiles allow us to
estimate the total mass of the cluster, assuming hydrostatic
equilibrium,
\begin{equation}\label{eqn:mtot}
M_\mathrm{HSE}(<r)=-\frac{rT_\mathrm{3D}(r)}{\mu m_p
  G}\left(\frac{\mathrm{dln}\rho_g}{\mathrm{dln}r}
  +\frac{\mathrm{dln}T_\mathrm{3D}}{\mathrm{dln}r}\right).
\end{equation}
where $\mu\approx 0.59$ is mean molecular weight for the fully ionized ICM, $m_p$ is the proton mass,
and we have set the Boltzmann constant to be unity such that the temperature $T_\mathrm{3D}(r)$
is in unit of energy. 


\section{Results}
\label{sec:results}


To test the MEKAL modeling of bremsstrahlung and metal line emissions
in low density regions, we perform the {\tt XSPEC} fit with a fixed
metal abundance parameter: $Z_\mathrm{fix}=0.5 Z_{\odot}$ for the
${Z}_{0.5}$ map, and $Z_\mathrm{fix}=0.05 Z_{\odot}$ for the
${Z}_{0.05}$ map.  The metallicity in the $Z_{0.05}$ map is close to
the simulation metallicity in cluster outskirts, but we use the
$Z_{0.5}$ map to also check the effects of a possible scenario where a
higher abundance of metals sits in cluster outskirts.  We then compare
results from each ${Z}_\mathrm{fix}$ case with the ${Z}_\mathrm{fit}$
case, where we perform the fit for all three parameters.

To eliminate the effects of a multi-temperature medium, we also
perform the {\tt XSPEC} fit on our $Z_{0.5}T_{X,1\text{ keV}}$ map.
We generate ICM profiles using a the best-fit metal abundance
parameter and fixed temperature, and compare with the profiles
resulting from the known input metal abundance and temperature value.

\subsection{Recovered temperature and density profiles with known
  abundance measurements}\label{sec:known_abundance}

In this section, we check for consistency in gas density measurements
from spectral fitting with the XSPEC fit performed using only two free
parameters: projected X-ray temperature and normalization of fit.  We
use the $Z_{0.5}$ and $Z_{0.05}$ photon maps for this experiment, and
fix the abundance parameter to the true abundance of the map.

\begin{figure*} 
  \centering
  \mbox{\subfigure{\includegraphics[width=3.5in]{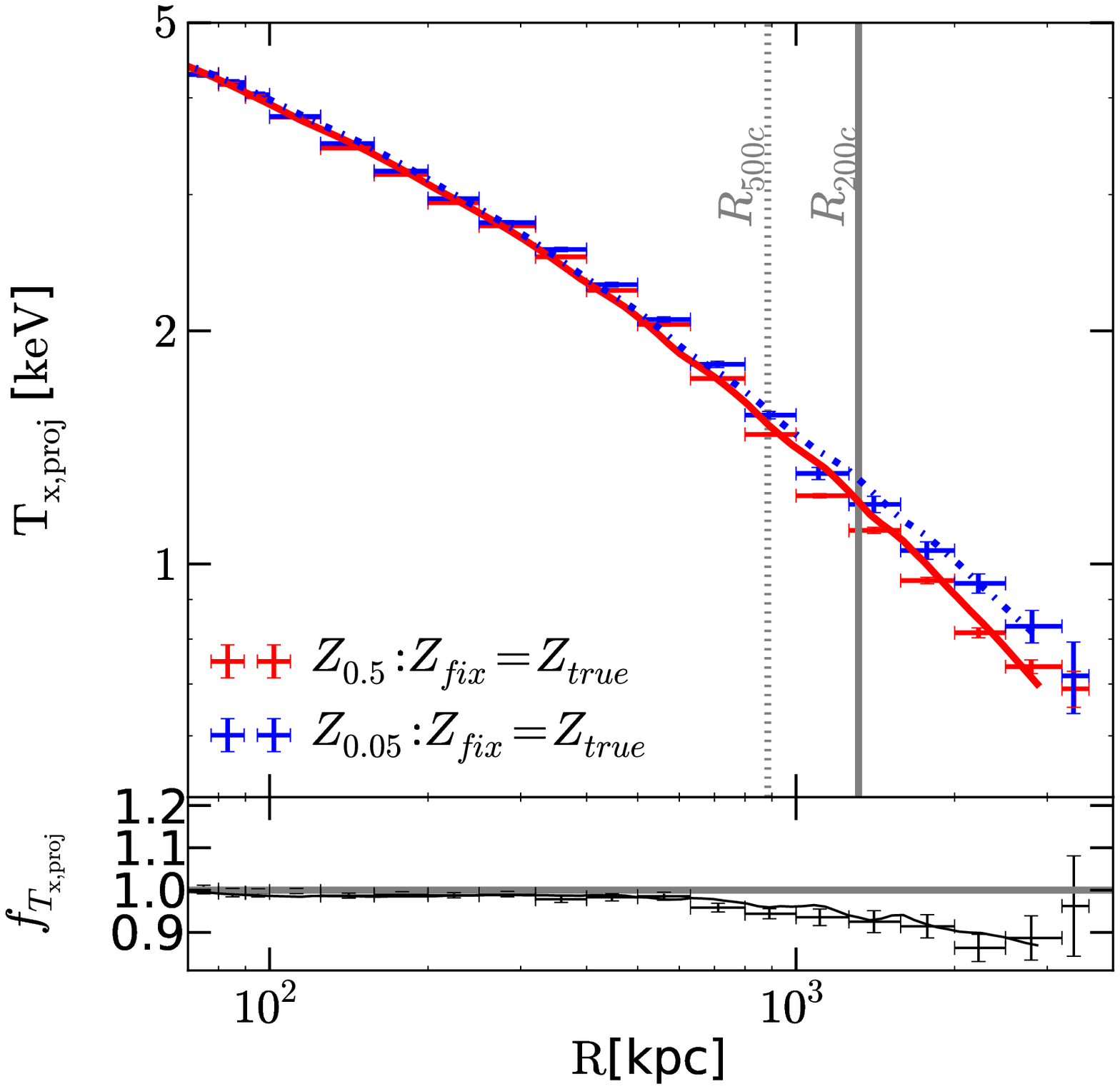}}\quad
    \subfigure{\includegraphics[width=3.5in]{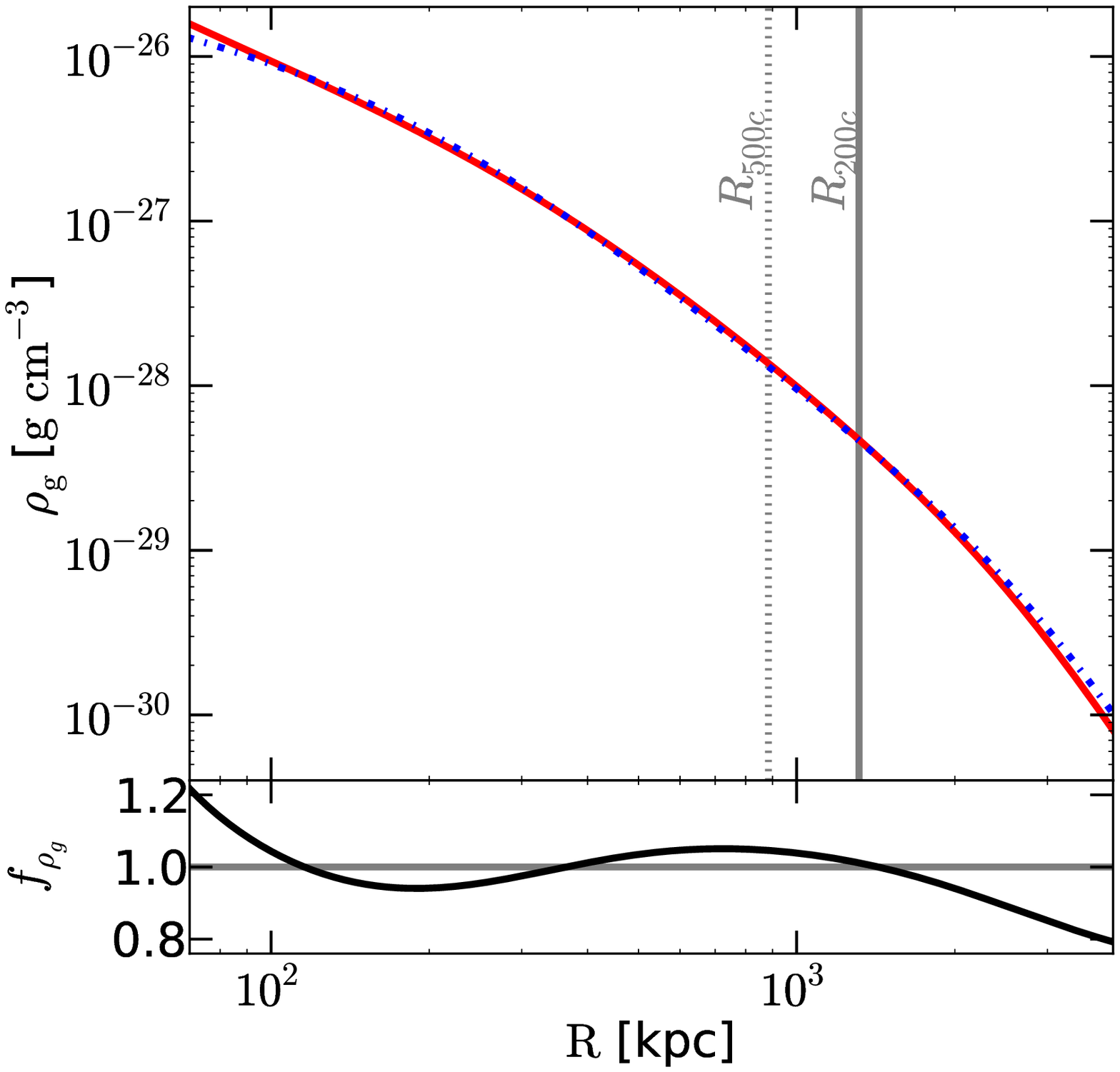}} }
  \caption{(Left) The projected X-ray temperature from spectral
    fitting and (right) the corresponding deprojected 3-D density
    profile for CL7.  Spectral fits used a fixed abundance parameter
    set to the true abundance of the photon map.  Mock maps were
    generated with constant fixed metallicity
    (${Z}_{0.5}=0.5{Z}_{\odot}$ in red solid and
    ${Z}_{0.05}=0.05{Z}_{\odot}$ in blue dash-dotted). The black line
    in the $f_{{T}_{X}}$ and $f_{\rho_g}$ axes is the ratio between
    the profile obtained using a fixed ${Z}_{0.5}$ and the profile
    using a fixed ${Z}_{0.05}$.  The black circles indicate the ratio
    between the $T_{X,\mathrm{proj}}$ data points.  Spectral fitting
    of two maps with identical gas density and temperature, but
    differing abundances, yield recovered density profiles that differ
    by up to $20\%$.}
   \label{fig:fix_abundance}
\end{figure*}

\begin{figure} 
  \centering
  \includegraphics[width=3.5in]{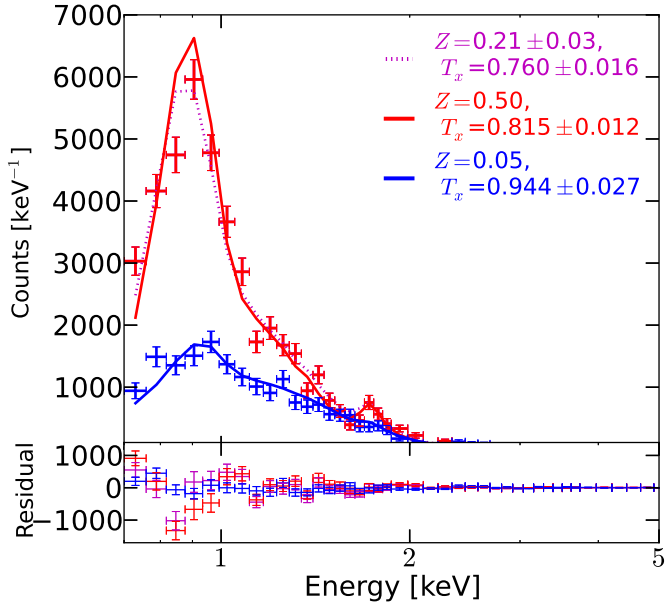}
  \caption{Energy spectra from the $Z_{0.5}$ (red) and $Z_{0.05}$
    photon maps corresponding to the third to the last radial bin in
    Figure~\ref{fig:fix_abundance} ($R\approx2\times{R}_{200c}$).
    Spectra are normalized to the size of the energy bin.  The solid
    lines show the best fit model spectra with the abundance parameter
    fixed to the true abundance of the photon map.  The dotted line
    corresponds to the best fit model spectrum allowing both the
    abundance and temperature values to vary. The legend labels the
    corresponding metallicity and best fit projected X-ray
    temperature.  We show the residual value between the model and our
    mock data in the lower panel. The statistically significant
      difference between all of the best fit X-ray temperature values
      is due to a degeneracy in fitting a multi-temperature medium
      with a single temperature fit.  The best fit abundance parameter
      for the $Z_{0.5}$ map in this bin is lower than the true
      abundance to compensate for the temperature that has been biased
      low.  Note, the lower metal abundance in the $Z_{0.05}$ map
    produces fewer photons from metal line cooling, leading to
    suppressed spectral features.  There are $18,800$ counts in the
    bin from the $Z_{0.05}$ map, and $43,000$ counts in the bin from
    the $Z_{0.5}$ map.}
   \label{fig:spectra_fixed}
\end{figure}

Figure~\ref{fig:fix_abundance} shows the best-fit value for the
projected X-ray temperature for each map.  In cluster outskirts, the
best fit X-ray temperature for the $Z_{0.05}$ maps is systematically
higher than that of the $Z_{0.5}$ map, and the gas density differs by
$\sim 20\%$.  Note that while the {\em true} temperature of the two
maps in each radial bin is identical, their measurements differ up to
$\sim 15\%$.  The discrepancy points to the degeneracies between
fitting parameters.  While the fitting procedure assumes a
single-temperature fit, the true ICM is a multi-temperature medium,
thus requires more degrees of freedom to capture the inhomogeneities.
Both the metal abundance and the normalization of the fit offer
additional degrees of freedom to minimize the residuals, so it is not
surprising that statistically significant differences can arise in
measured X-ray temperatures if the metal line contributions are
different.  We plot the spectra of photons from the radial bin with
the largest deviation in measured X-ray temperature $(R\approx2\times
R_{200c})$ in Figure~\ref{fig:spectra_fixed}.  Compared to the
spectrum from the $Z_{0.5}$ photon map (red points), the spectrum from
the $Z_{0.05}$ map (blue points) has a much lower emission from metal
line cooling, and therefore it has a suppressed peak in the energy
spectrum.  The solid lines show the best fit model spectrum for each
map.

\begin{figure} 
  \centering
  \includegraphics[width=3.5in]{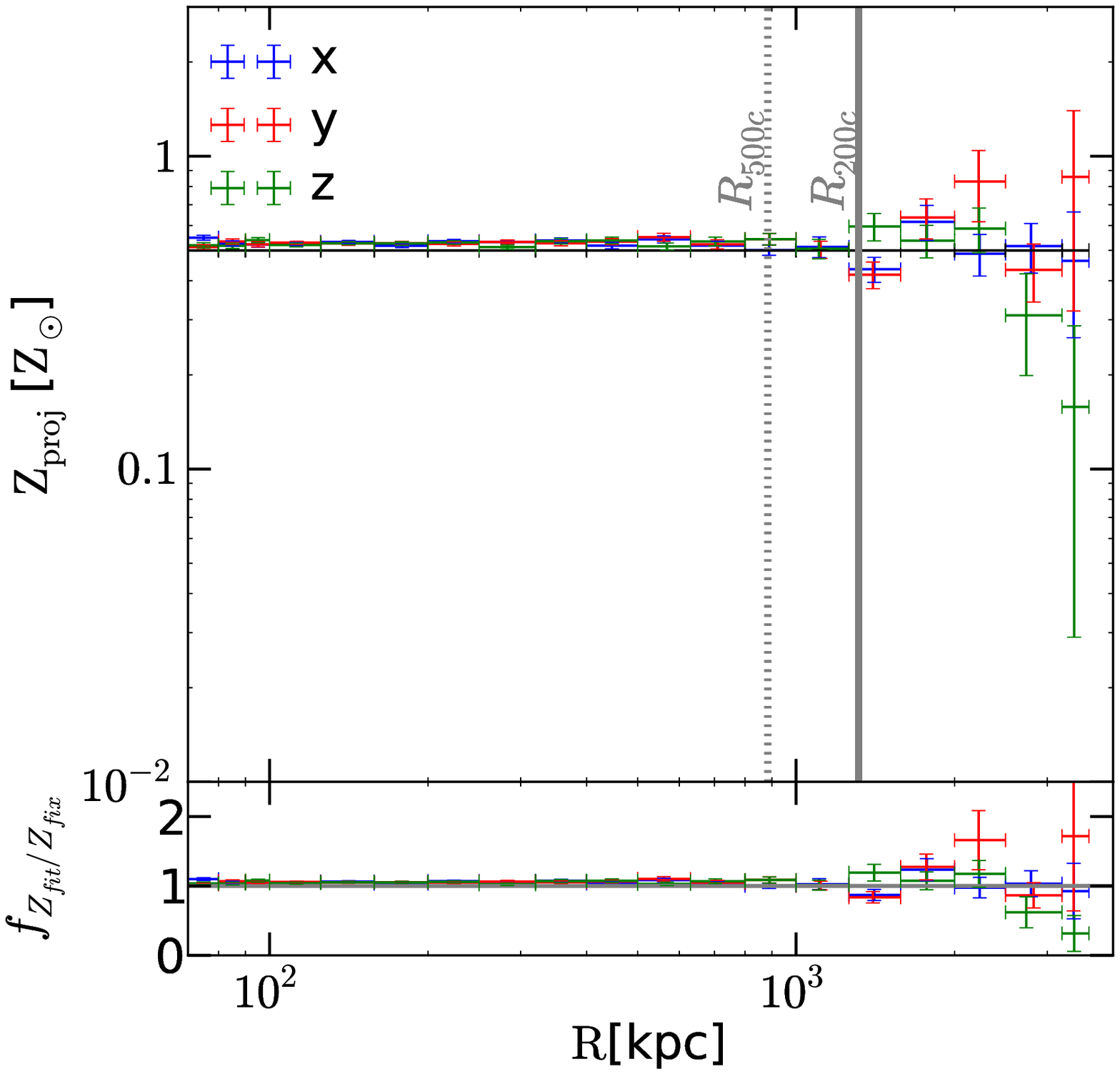}
  \caption{The best-fit projected metallicity along three
      different projections of CL7. Mock maps were generated with both
      a constant fixed metallicity of ${Z}_{0.5}=0.5{Z}_{\odot}$ and
      constant fixed temperature of $T_X=1$ keV.  The lower axis shows
      the ratio between the best-fit metal abundance and the true
      metal abundance. X-ray temperature and normalization are free
      parameters.  In the absence of a multi-temperature medium, the
      abundance is well recovered out to large radii.}
   \label{fig:known_temperature}
\end{figure}

With a metal abundance parameter fixed to the true value of
$Z=0.5\text{Z}_\odot$, the model spectrum overshoots the peak of the
mock photon spectrum at the Fe-L complex between 0.8 and 1.4 keV, and
the best fit X-ray temperature is $T_X=0.81\pm0.01$keV.  If we allow
the metal abundance to vary as an additional free parameter, both the
best fit metal abundance ($Z=0.21\pm0.03\text{Z}_\odot$) and the X-ray
temperature ($T_X=0.76\pm0.02$keV) are lower and the residual
difference between the mock photon counts and the model spectrum
improves.

In Figure~\ref{fig:known_temperature}, we show how the systematic bias
in metal abundance goes away in the absence of a multi-temperature
ICM.  The axes show the best fit metal abundance as a function of
radius along three different lines of sight for our
$Z_{0.5}T_{X,1\text{ keV}}$ map.  This test isolates biases due to
density inhomogeneities, lower photon counts, and instrumental
response.  For a map with uniform temperature, the fitting procedure
well recovers the input constant metallicity, and does not exhibit a
systematic decline.  The best fit projected X-ray temperature is also
well recovered to the fixed input value.

The suppressed peak in the mock photon spectrum in
Figure~\ref{fig:spectra_fixed} can be attributed to a high temperature
phase of gas in the outskirts.  For gas with an average temperature of
$T_X\approx1$ keV, the Fe-L line would be sensitive to a higher
temperature component of gas, even if the metallicity of that
component is the same (e.g., see Figure~3 in \citet{mazzotta_etal04}
and Figure~23 in \citet{bohringerandwerner_10}).  

Note, if the abundance parameter were not fixed (magenta dotted line
in Figure~\ref{fig:spectra_fixed}, the best fit abundance parameter
would have to be lower than the true value $Z=0.5Z_{\odot}$ to better
minimize the residuals in the fitting procedure.  If we were to
increase the X-ray temperature in the model spectrum, the residual at
the peak would improve, but at the expense of increasing the residual
values at other energies.

On the other hand, for the $Z_{0.05}$ spectra in
Figure~\ref{fig:spectra_fixed}, there are fewer photons from metal
line cooling.  Hence, most of the spectral features are hidden in
Poisson noise.  Because of the suppressed spectral features, the
fitting procedure produces a best-fit X-ray temperature and
normalization that will better minimize the residuals near the peak of
the spectrum as opposed to at energies with much fewer counts.  To
minimize the residuals near the peak, the fitted X-ray temperature
must then be higher to harden the model spectrum, which is why the
best fit projected X-ray temperature for the $Z_{0.05}$ map is
systematically higher in the outskirts, where photon counts are low.

\subsection{Testing metal abundance measurements in cluster
  outskirts}\label{sec:test_model_spectra}
The robustness of metal abundance measurements in the low density
outskirts is not well understood.  In this section, we compare the
abundance measurements from spectral fitting to the true abundance
from our photon maps.  We also discuss the effects on the projected
X-ray temperature, projected emission measure and the reconstructed
three dimensional gas density profile.  We begin by comparing
variations in projected X-ray temperature $T_{X,\mathrm{proj}}$ and
abundance measurements $Z_{\mathrm{proj}}$ from our ${Z}_{0.5}$ and
${Z}_{0.05}$ photon maps, where we have set the input metal abundance
to constant values.

\begin{figure*} 
  \centering
  \mbox{
    \subfigure{\includegraphics[width=3.5in]{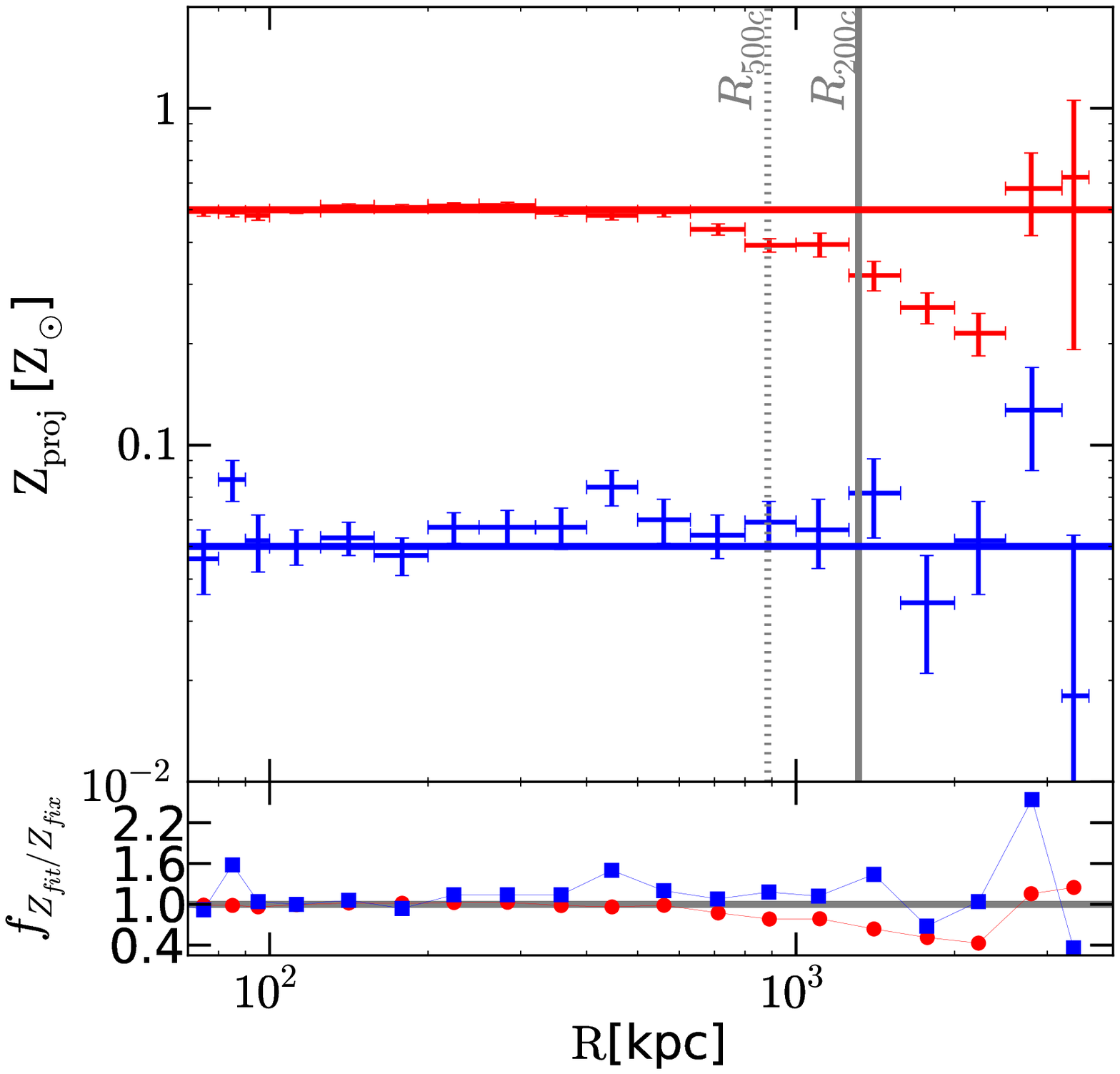}}\quad
    \subfigure{\includegraphics[width=3.5in]{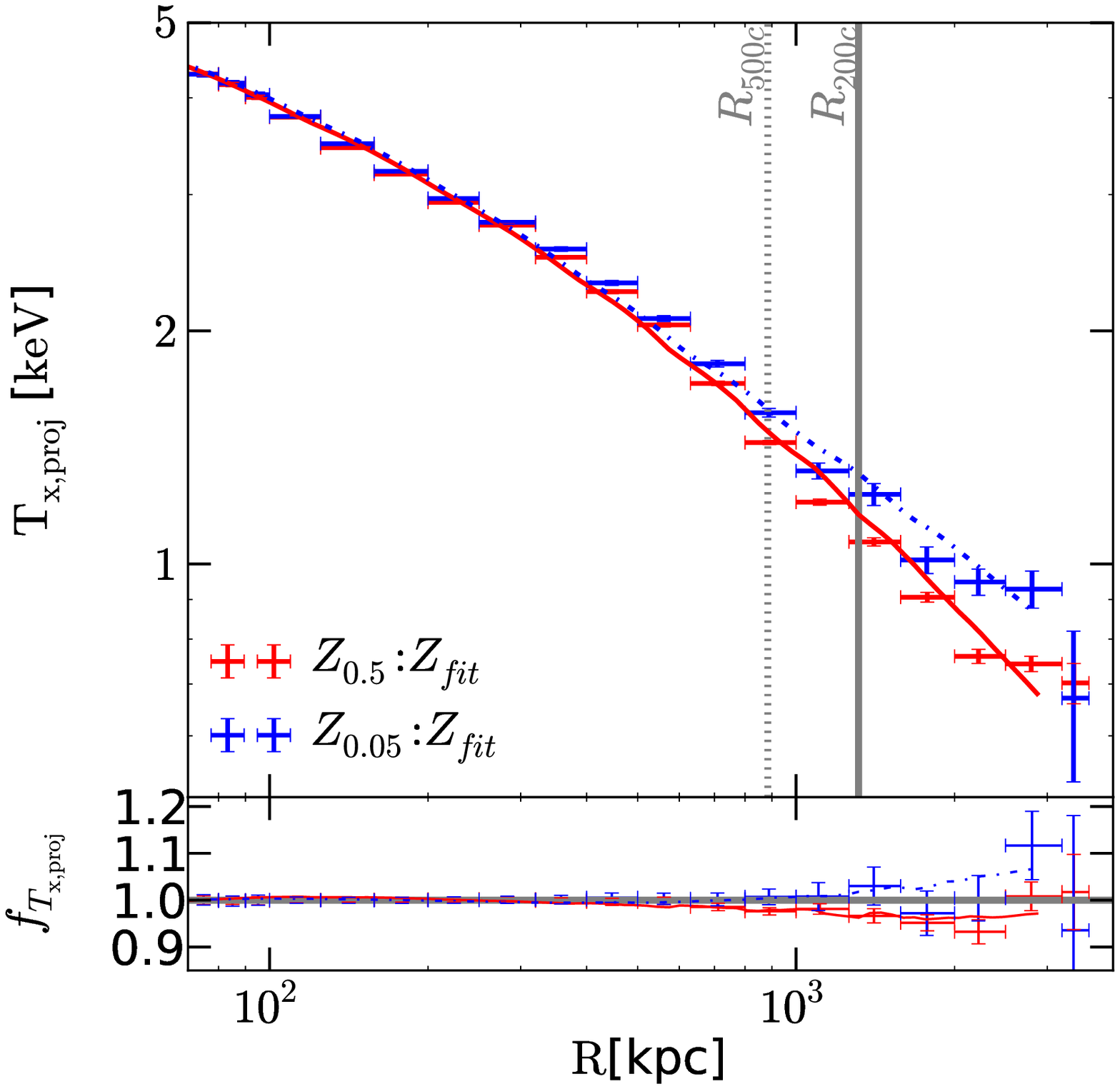}}
  }

  \mbox{
    \subfigure{\includegraphics[width=3.5in]{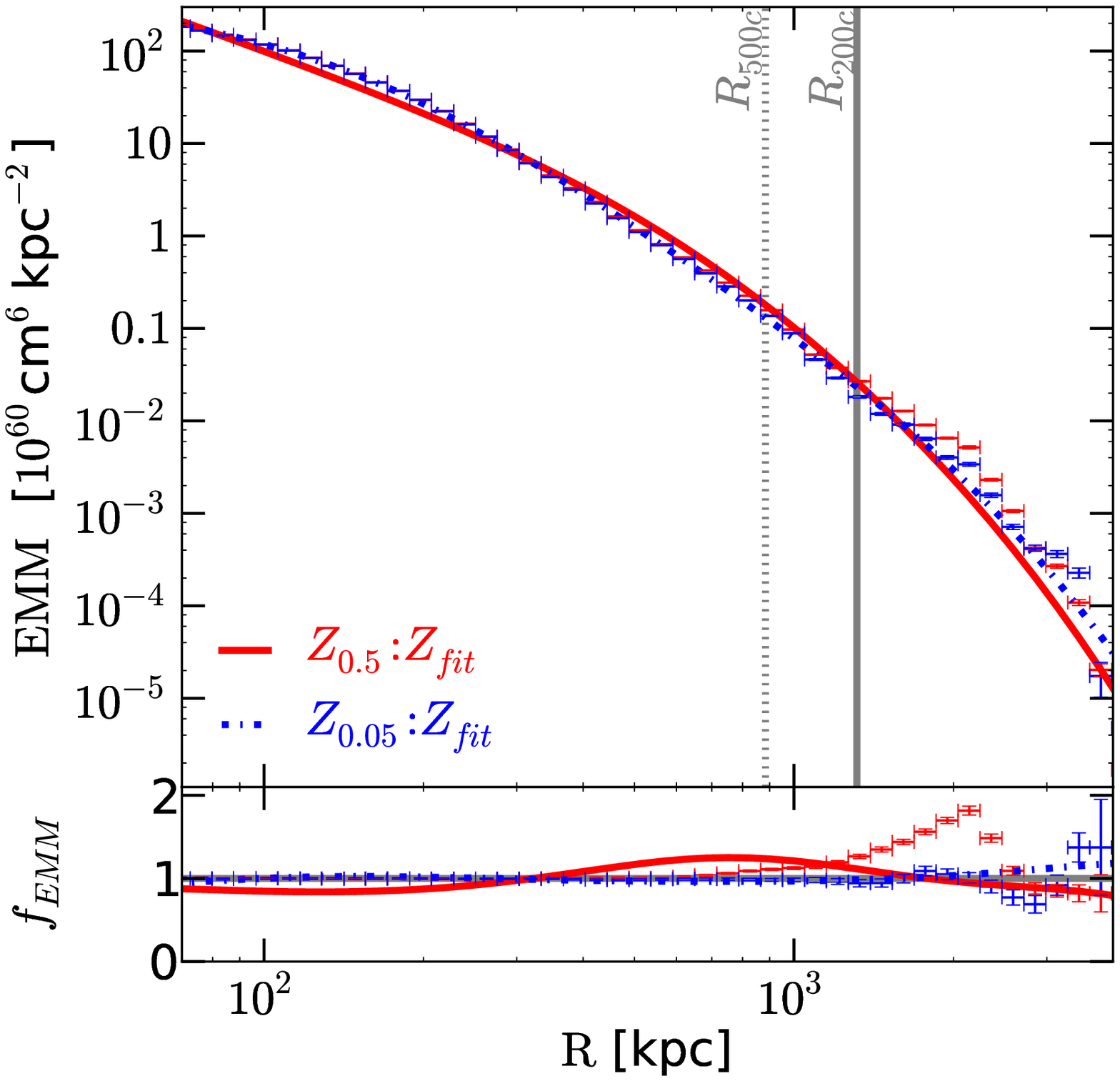}}\quad
    \subfigure{\includegraphics[width=3.5in]{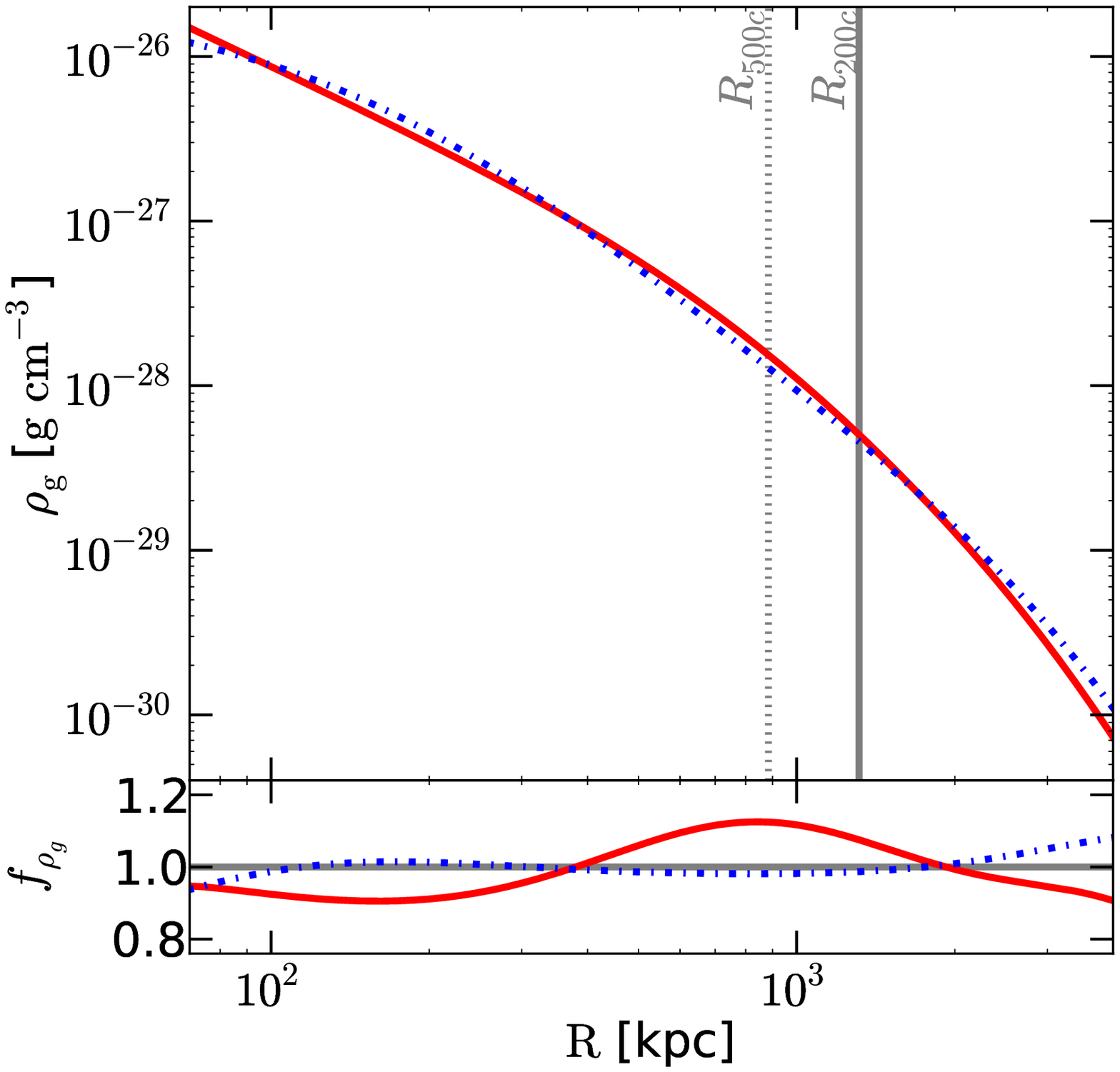}} 
  }
  \caption{From top left panel in a clockwise direction, the projected
    metallicity, projected X-ray temperature, recovered 3-D density
    profiles, and projected emission measure for CL7. Mock maps were
    generated with constant fixed metallicity
    (${Z}_{0.5}=0.5{Z}_{\odot}$ in red solid and
    ${Z}_{0.05}=0.05{Z}_{\odot}$ in blue dash-dotted).  The lower
      axes, labeled with an $f$, show the ratio between profiles
    recovered with metal abundance as a free parameter to profiles
    recovered with abundance fixed to the known input abundance.  The
    systematic decline of the best-fit projected metallicity beyond
    ${R}_{500}$ is due to the multi-temperature medium, as shown in
    Figure~\ref{fig:spectra_fixed}. The effects of low photon count
    from metal line cooling is visible in both the ${Z}_{0.05}$ best
    fit projected metallicity as well as in the last two bins of
    ${Z}_{0.5}$.  The bias in metal abundance measurement introduces
    variations of up to $15\%$ in recovered density profiles.}
   \label{fig:fitfix_abundance}
\end{figure*}

Figure~\ref{fig:fitfix_abundance} compares ICM profiles from our
$Z_{0.5}$ and $Z_{0.05}$ mock X-ray flux maps of CL7, projected along
the $z$-axis. Results along other axis projections are qualitatively
similar.  The top left panel shows the metal abundance measurements
retrieved from our mock spectra, fitting the mock spectra to MEKAL
model spectra using XSPEC.  We define
$f_{Z_\mathrm{fit}/Z_\mathrm{fix}}$ as the ratio between the best fit
abundance parameter to the true abundance of our photon map. The
$f_{Z_\mathrm{fit}/Z_\mathrm{fix}}$ profile shows that the while XSPEC
fitting procedure recovers a $Z_\mathrm{fit}$ value that is within a
factor of 2 of the true input abundance of the photon map, the
abundance recovery is significantly worse at larger radii.

For the $Z_{0.5}$ map, the abundance measurements has a systematic
decline from the true value at large radii $R\gsim 0.7 R_{200c}$.  As
discussed in Section~\ref{sec:known_abundance}, {\it the metal
  abundance parameter adjusts to compensate for the multi-temperature
  medium when we perform a single-temperture spectral fit}.  
Thus, the best fit abundance parameter will be smaller
than the true abundance in order to minimize the residual around the
suppressed emission lines.  However, if there are too few photons from
metal line cooling, the poisson errors become large, and therefore
this systematic effect is less apparent since the statistical errors
dominate. This is why the last two radial bins do not show
underestimation in abundance measurements.

For the $Z_{0.05}$ map, the photon counts around the peak of the
spectrum ($0.7-1$~keV) is a factor of three lower than that of the
$Z_{0.5}$ spectrum (See Figure~\ref{fig:spectra_fixed}).  In addition
to the intrinsically less pronounced metal lines, the low photon count
makes it more difficult for the fitting routine to discern between
contributions from metal line cooling and thermal bremsstrahlung
emissions, leading to more scatter in the abundance measurements shown
in Figure~\ref{fig:fitfix_abundance}.  Note that similar scatter is
seen in the last two radial bins from the $Z_{0.5}$ map, which also
has a very low photon count.

The top right panel of Figure~\ref{fig:fitfix_abundance} shows the
projected X-ray temperature profile (red and blue points for the
respective abundances) from spectral fitting, and the corresponding
best fit projected 3D temperature model (thin lines of the same
color).  We measure the relative bias in $T_{X,\mathrm{proj}}$ as the
ratio of the projected X-ray temperature measured with abundance as a
free parameter (the profile in Figure~\ref{fig:fitfix_abundance}
compared with the X-ray temperature measured with the abundance
parameter fixed to the true abundance of the map (see
Figure~\ref{fig:fix_abundance}).  The X-ray temperature measurements
from fits with abundance as a free parameter ($Z=Z_\mathrm{fit}$) vary
up to $\sim10\%$ of the measurements with a fixed abundance parameter
($Z=Z_\mathrm{true}$).  Note, the relative bias in projected X-ray
temperature from spectral fitting is in the same direction as that of
the abundance measurement.

\begin{figure} 
  \centering
  \includegraphics[width=3.5in]{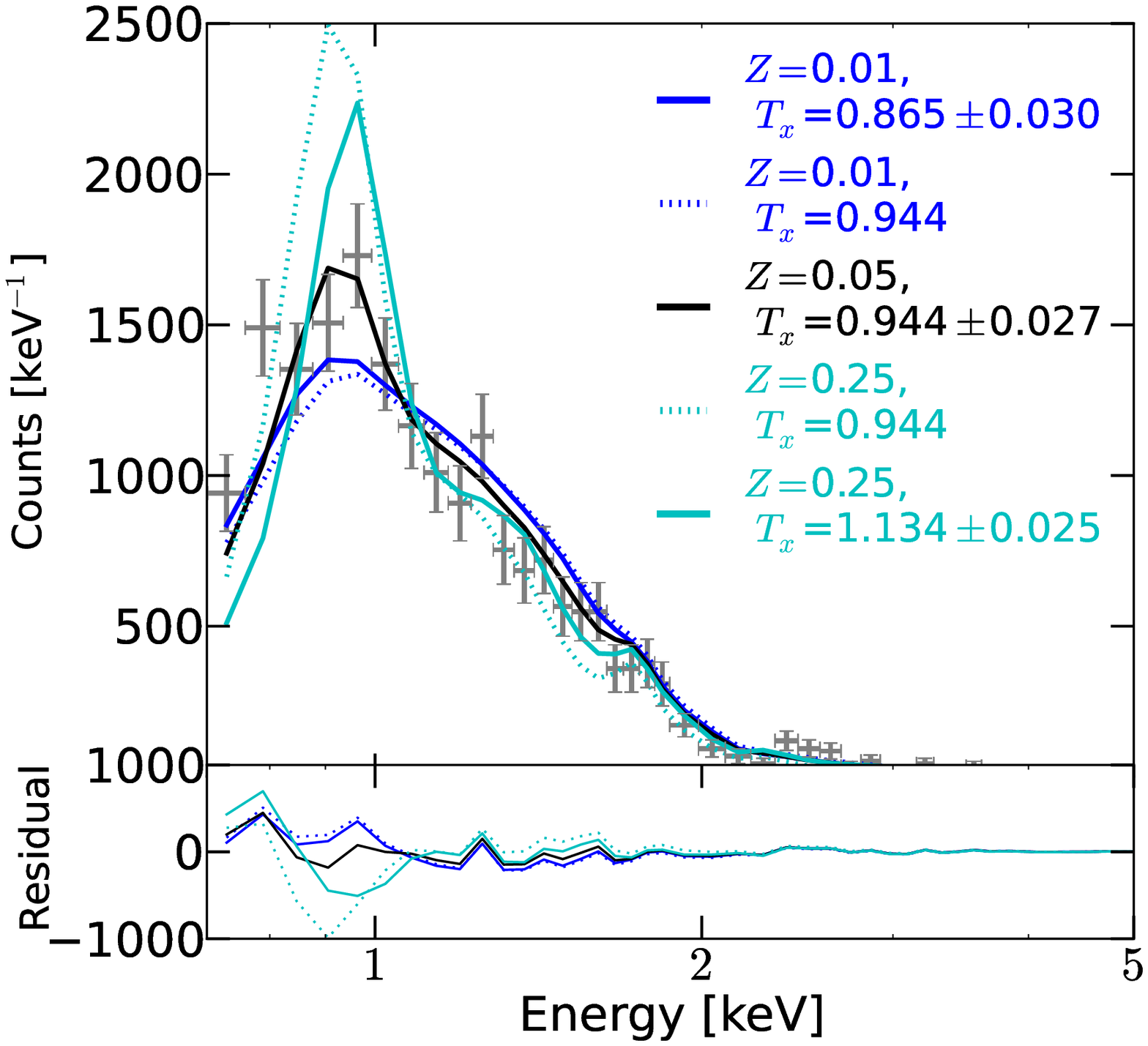}
  \caption{Grey errorbars show the energy spectrum from the $Z_{0.05}$
    photon map corresponding to the third to the last radial bin in
    Figure~\ref{fig:fix_abundance} ($R\approx2\times{R}_{200c}$).
    Spectra are normalized to the size of the energy bin.  The solid
    lines show the best fit model spectra with the abundance parameter
    fixed ($0.01Z_{\odot}$ in blue, $0.05Z_{\odot}$ in black, and
    $0.25Z_{\odot}$ in cyan), and the best fit projected X-ray
    temperature.  The dotted lines correspond to model spectra with
    the projected X-ray temperature fixed to
    $T_{X,\mathrm{proj}}=0.944$ keV, which is the value of the
    best-fit X-ray temperature with $Z_\mathrm{fix}=Z_\mathrm{true}$.
    We show the residual value between the model and our mock data in
    the lower panel.  The model spectra show that the best-fit
    projected X-ray temperature will be biased in the same direction
    as the best-fit abundance parameter, as shown in
    Figure~\ref{fig:fitfix_abundance}.}
   \label{fig:spectra_fixed_highlow}
\end{figure}

The cause of the correlation in the relative bias of both the
abundance and X-ray temperature becomes evident when we compare model
spectra (see Figure~\ref{fig:spectra_fixed_highlow}).  When the best
fit abundance is higher (lower) than the true abundance, the peak of
the model spectrum will sit above (below) the mock spectrum, and a
higher (lower) X-ray temperature is necessary to harden (soften) the
spectrum to minimize the residuals.  Note, while fixing the metal
abundance parameter to a value that is five times the true value
significantly biases the value of the best fit X-ray temperature
parameter, it does not significantly increase the error in the X-ray
temperature parameter.  The X-ray temperature fit is mostly sensitive
to shape of the continuum, and most of the residual contributions to
the error in X-ray temperature come from energy bins outside of the
Fe-L complex. Here, the model spectra are still within the errorbars
of the mock data.

An interesting feature to note in the projected X-ray temperature
profile is that the projected 3D temperature model fits the data
points well only out to $R\lsim{R}_{200c}$.  While the best-fit model
profile is within the errorbars of the mock data, it does not capture
the emerging trend of the decreasing slope of the projected X-ray
temperature at larger radii.  The departure from the smooth
temperature model profile is indicative of the complicated,
inhomogeneous temperature structure in cluster outskirts. The smooth
temperature model does not have enough degrees of freedom to account
for the bumpier features of the $T_{X,\mathrm{proj}}$ data points,
indicating that perhaps a modified fitting formula would be
appropriate for total mass measurements outside of ${R}_{200c}$.

Note, the errorbars at large radii for the metal abundance
measurements are not as good as those of the X-ray temperature
measurements because the abundance fitting requires an order of
magnitude more photons to have the same statistical uncertainty as the
X-ray temperature.  The metal abundance is only affected by line
emission contributions as opposed to the X-ray temperature, which also
has contributions from thermal bremsstrahlung.

From Equation~(\ref{eqn:cnt2emm}), we can see that for the same X-ray
surface brightness $XSB$ at a given radius, a higher (lower) gas density is
necessary to compensate for a lower (higher) X-ray emissivity
$\Lambda$.  The bottom two panels in Figure~\ref{fig:fitfix_abundance}
show the corresponding projected emission measure and gas density for
the same photon maps as in the top two panels.  The projected
emission measure varies to $20\%$ at all radii, and the gas density to
a level of $15\%$.  

We note that the apparent enhanced bias in projected emission measure
at intermediate radii for the ${Z}_{0.5}$ map, shown in red, is due to
effects from integrating over the line of sight.  The density profile
measured with a fixed abundance parameter is steeper at intermediate
radii than the density profile measured with the best fit values for
the abundance parameter.  Hence, along an equal line of sight, there
is a relatively lower emission measure in the fixed abundance case 
than in the best fit abundance case.

In summary, the relative bias in abundance and projected X-ray
temperature are correlated, and vary up to $50\%$ and $10\%$,
respectively.  Both the resulting projected emission measure and gas
density profiles vary to $10\%$ at low and intermediate radii, but
deviate as much as $40\%$ and $20\%$ in emission measure and gas
density, respectively.

\subsection{Systematic effects of metal abundance measurements on gas profile measurements}

As described in Section~\ref{sec:test_model_spectra}, a high (low) bias
of the abundance and X-ray temperature measurements would lead to a
suppressed (enhanced) density measurement in that radial bin.

In this section, we examine the systematic effects on gas profile measurements 
with a factor of 2 and 5 bias in the abundance fitting at all radii.  For this test, 
we use the $Z_{0.05}$ map where we take the metallicity to be constant at 
$Z=0.05Z_\odot$ throughout the simulation.
We choose this value since it is closer to the metal abundance 
at large radii ${R}\gtrsim{R}_{200c}$ in our self-consistent cluster simulation 
with metal enrichment and advection. 
We run XSPEC fitting by fixing the abundance parameter at metallicities 
a factor of 2 and 5 from the true value from the map:
$Z_\mathrm{fix}/Z_{\odot}=0.01, 0.025, 0.1, 0.25$, and we compare the results to those 
fitted with the true value $Z=0.05Z_\odot$. 
To isolate the effects from all other quantities, we only allow 
the projected X-ray temperature and the spectral normalization
to be free parameters in the fitting . 

\begin{figure*} 
  \centering
  \mbox{                                                                                                                            
    \subfigure{\includegraphics[width=3.5in]{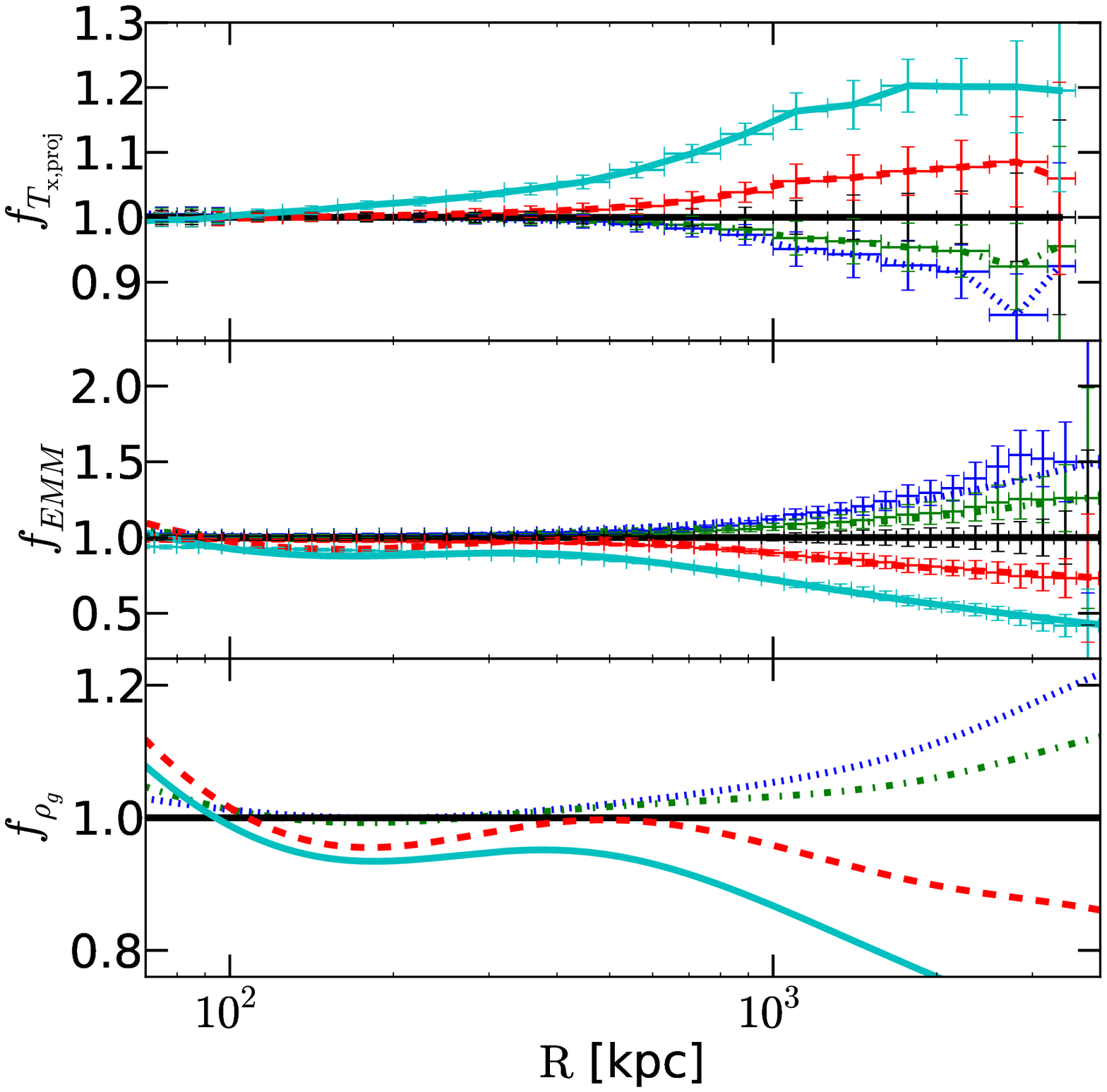}}\quad                                    
    \subfigure{\includegraphics[width=3.5in]{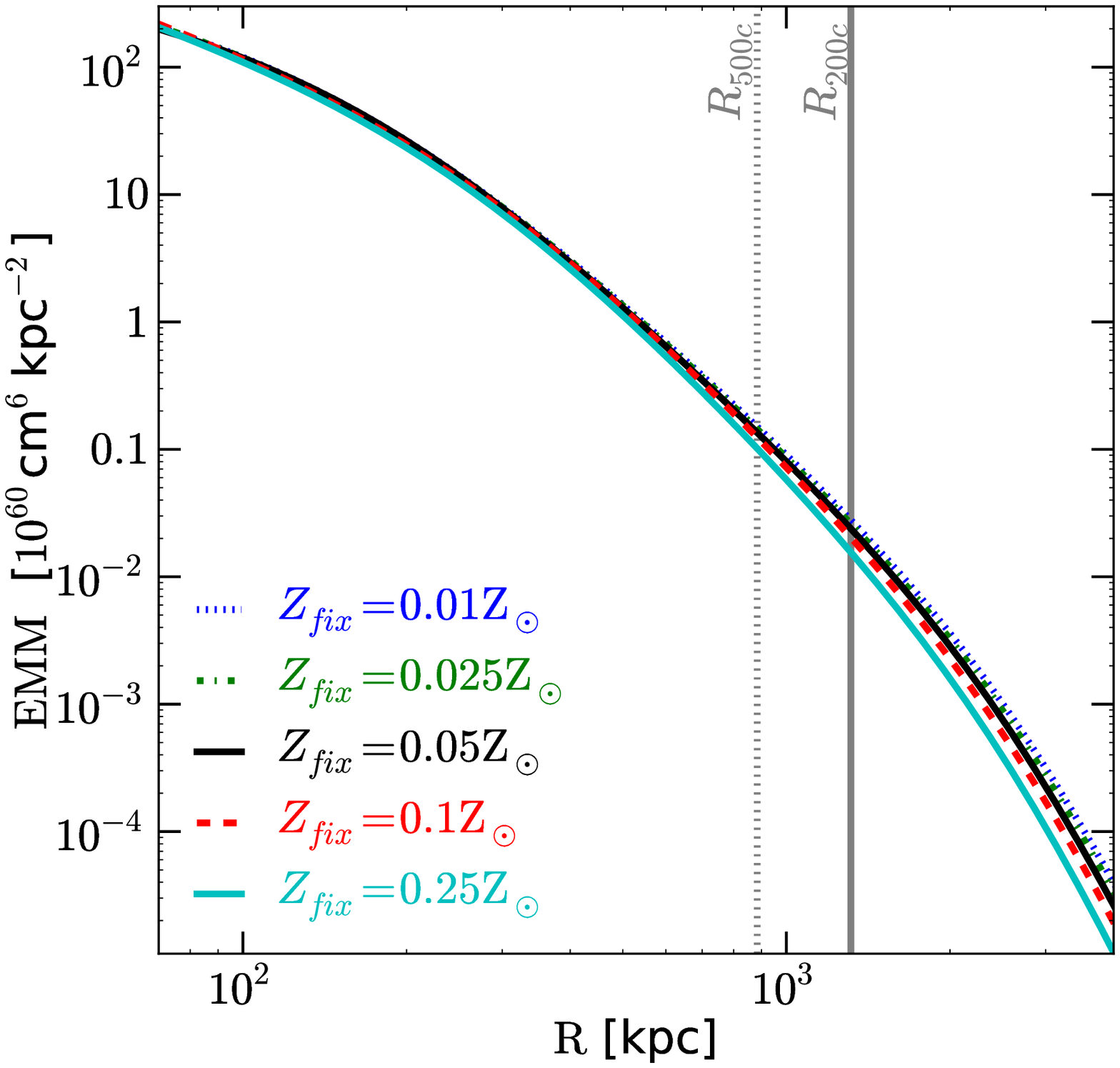}}                                        
  }
  \caption{The above figures are profiles corresponding to the
    ${Z}_{0.05}$ flux maps. The plot on the left shows the ratio
    between profiles using various values for ${Z}_\mathrm{fix}$, and
    the corresponding profile with
    ${Z}_\mathrm{fix}={Z}_{true}=0.05\mathrm{Z}_{\odot}$. From top to
    bottom panel, these respectively are the ratios in X-ray
    temperature, gas density, and projected emission measure.  The
    figure on the right shows the projected emission measure for each
    value of ${Z}_\mathrm{fix}$.  At $3\times{R}_{200c}$, an abundance
    measurement that is biased by a factor of 2 results in gas
    densities that are biased by $15\%$.}
  \label{fig:fixed_abundances}
\end{figure*}

In Figure~\ref{fig:fixed_abundances}, we plotted the X-ray
temperature, projected emission measure, and gas density for each case
of ${Z}_\mathrm{fix}$ and compare to the
${Z}_\mathrm{fix}={Z}_\mathrm{true}=0.05{Z}_{\odot}$ case (black solid
line). Consistent with what we found in
Section~\ref{sec:test_model_spectra}, an over (under) estimate of the
abundance will lead to suppressed (enhanced) emission measure and
subsequently gas density measurements.

If the abundance is measured to within a factor of two of the true
metal abundance, the resulting density measurements in the outskirts
will be within $15\%$.  However, for the most extreme case of an
abundance measurement biased by a factor of 5, the density
measurements will have up to a $35\%$ bias.

The emission measure profile on the right panel of
Figure~\ref{fig:fixed_abundances} shows that an abundance measurement
that is higher than the true value can steepen the profile in cluster
outskirts.  For observational measurements that extrapolate the
abundance at large radii, this could introduce an even larger bias 
in the emission measure, density and temperature.  Our tests that
fix the abundance measurement to be a factor of two and five times the
true abundance show how severely this extrapolation can propagate to
inferred quantities, even at ${R}_{200c}$.  Most notably, for an
assumed abundance that is a factor of five higher than the true
abundance, the gas density is biased low by $\approx15\%$.

\subsection{Hydrostatic mass and gas mass fraction}

\begin{figure*} 
  \centering
  \mbox{                                                                                                                            
    \subfigure{\includegraphics[width=3.5in]{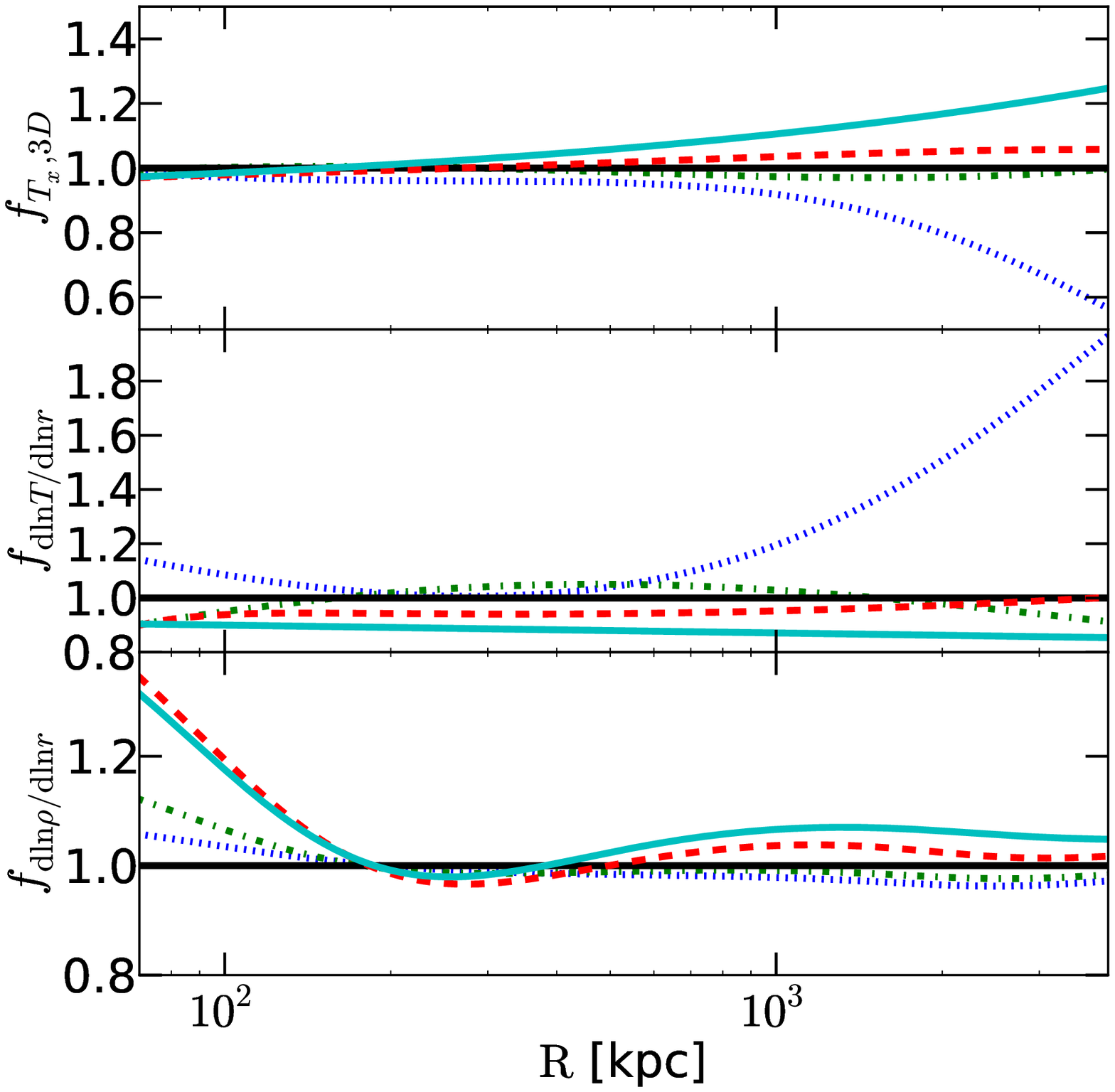}}\quad                                   
    \subfigure{\includegraphics[width=3.5in]{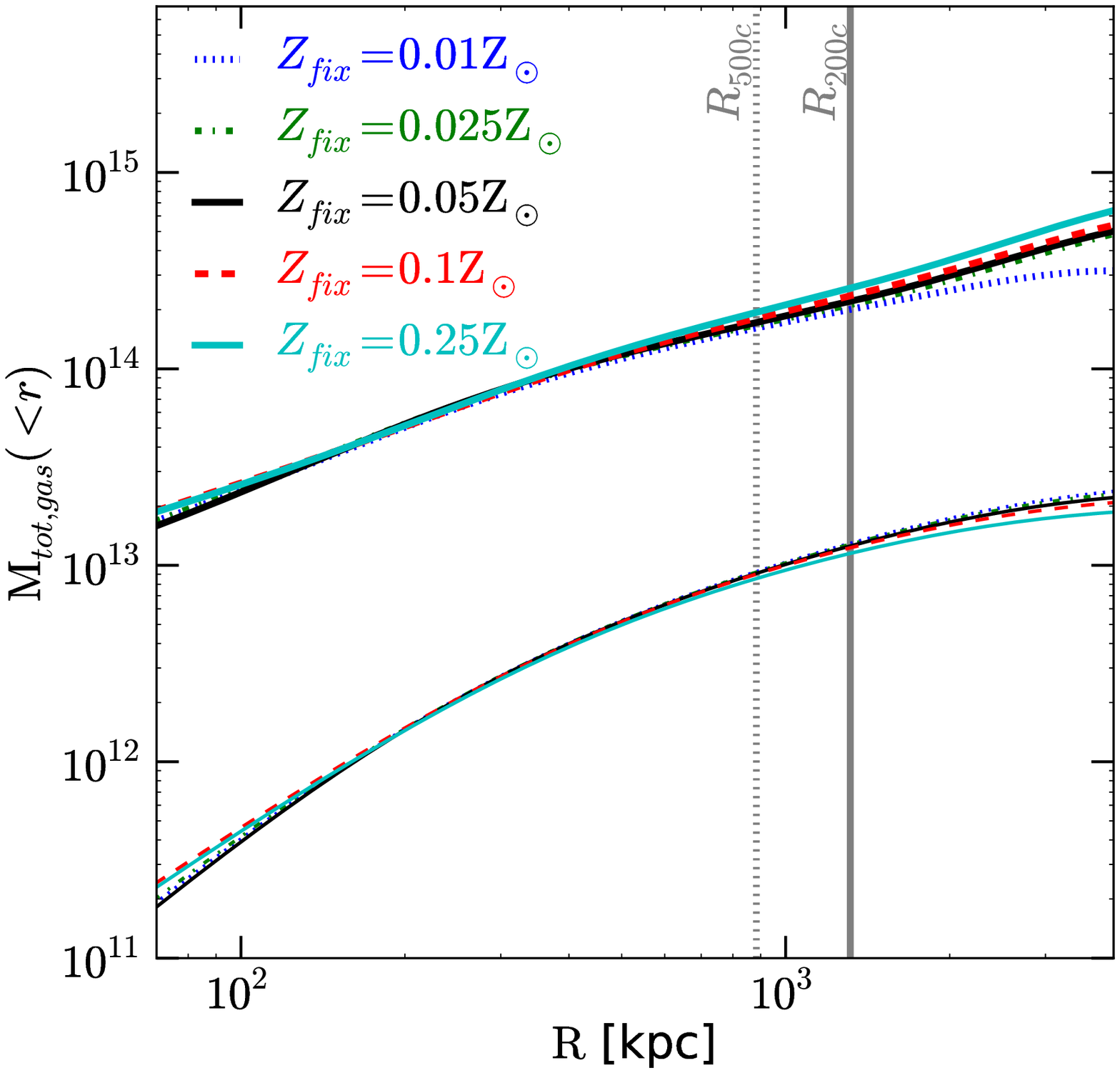}}                                        
  }
  \caption{The above figures are profiles corresponding to the
    ${Z}_{0.05}$ flux maps, with various values of ${Z}_\mathrm{fix}$,
    identical to the data set from Figure~\ref{fig:fixed_abundances}. Left:
    Ratios of quantities that contribute to the total mass
    measurement - the best fit 3D X-ray temperature, the logarithmic
    slope of the 3D temperature profile, and the logarithmic slope of the
    density profile. Right: The total mass profile (thick lines), and
    the gas mass profile (thin lines).  At $3\times{R}_{200c}$, an
    abundance measurement that is biased by a factor of 2 results a
    logarithmic gas density slope that is biased by $<5\%$.}
  \label{fig:fixed_abundances_masses}
\end{figure*}

From Equation~(\ref{eqn:mtot}), we can see that the X-ray measurement of the
hydrostatic mass of the cluster is sensitive to three quantities: the best
fit 3D model of the temperature profile, the logarithmic slope of
the 3D temperature profile, and the logarithmic slope of the gas
density profile. The left panel of
Figure~\ref{fig:fixed_abundances_masses} illustrates how each of these
quantities is affected by a systematic over (under) fit abundance.

\begin{figure*} 
  \centering
  \mbox{                                                                                                                            
    \subfigure{\includegraphics[width=3.5in]{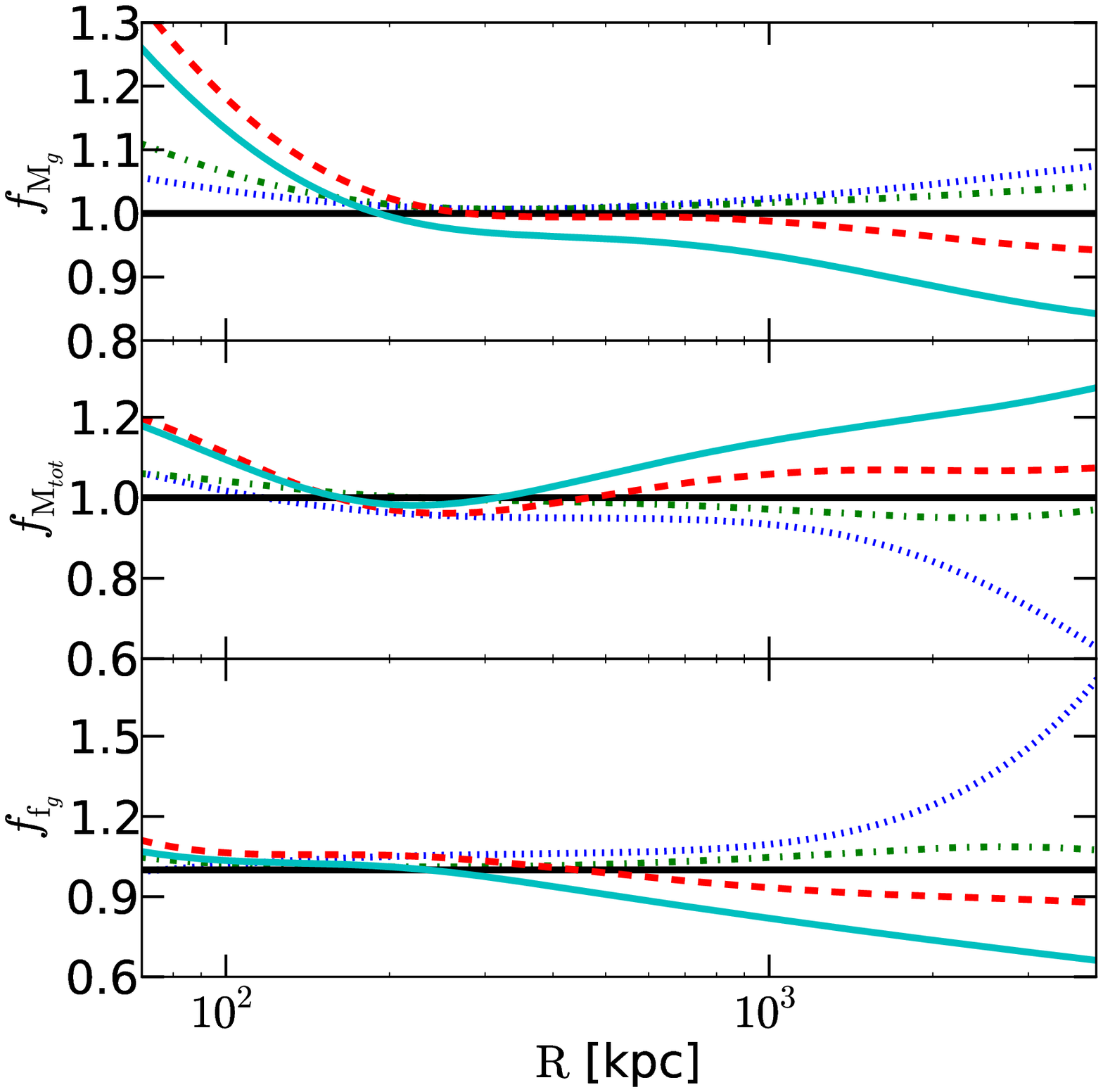}}\quad                               
    \subfigure{\includegraphics[width=3.5in]{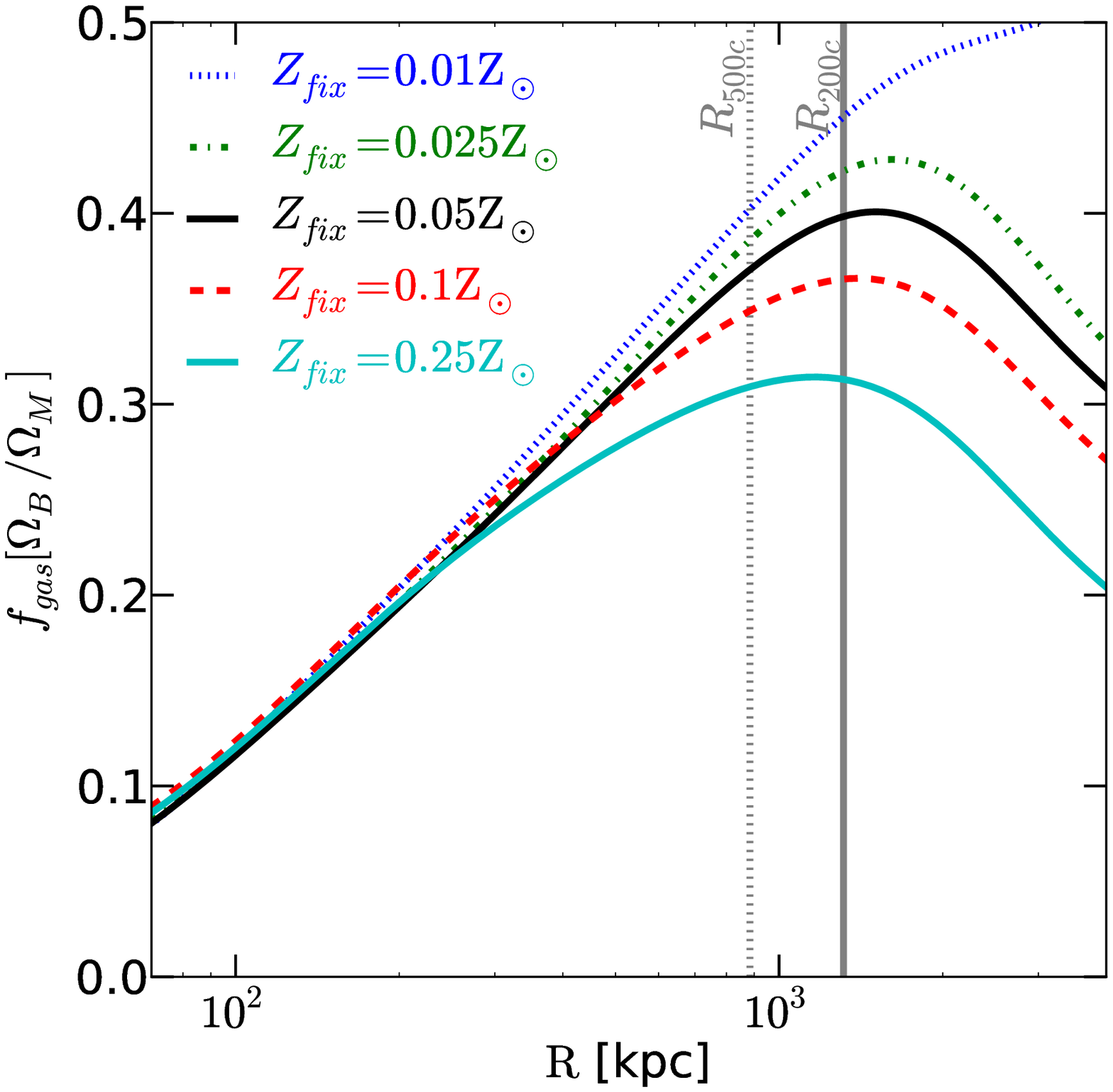}}                                        
  }
  \caption{The above figures are profiles corresponding to the
    ${Z}_{0.05}$ flux maps, with various values of ${Z}_\mathrm{fix}$,
    identical to the data set from
    Figure~\ref{fig:fixed_abundances}. The panels on the left are
    ratios of the gas mass, hydrostatic mass, and gas mass fraction, and
    the figure on the right is the gas mass fraction profile for each fixed
    value of the metal abundance.  At $3\times{R}_{200c}$, an abundance
    measurement that is biased by a factor of 2 results in a
    fractions that is biased by $\sim5\%$.}
  \label{fig:fixed_abundances_fgas}
\end{figure*}

For the case of recovering the metal abundance to within a typical
factor of two ($Z_\mathrm{fix}=0.025{Z}_{\odot}$ and
$Z_\mathrm{fix}=0.1{Z}_{\odot}$), all three quantities are within
$<10\%$ at intermediate and large radii.  At large radii, the bias in
the 3D temperature and logarithmic density slope dominate to drive the
hydrostatic mass bias in the same direction. A lower temperature fit
in the outskirts corresponds to a shallower logarithmic density slope,
and therefore a smaller hydrostatic mass (see dotted blue and
dot-dashed green thick lines in the right panel of
Figure~\ref{fig:fixed_abundances_masses}). However, the lower fitted
temperature and shallower logarithmic slope correspond to a larger
enclosed gas mass (see corresponding thin lines in
Figure~\ref{fig:fixed_abundances_masses}), as underestimates in both
abundance and temperature lead to overestimates in emission measure,
therefore bias the gas density high.

Figure~\ref{fig:fixed_abundances_fgas} shows that the combined bias in
enclosed gas mass and the hydrostatic mass drives the bias in gas
fraction up for the low fixed abundance values, and down for the high
fixed abundance values. This shows that a poor fit to the X-ray
temperature and metallicity in cluster outskirts is potentially
causing biases in the {\em Suzaku} gas mass fraction measurements at
large radii. The same magnitude of errors can come from an incorrectly
extrapolated metal abundance in cluster outskirts.

For a standard factor of 2 error in the abundance measurement, the gas
mass fraction is only affected by $10\%$, but any underestimate of the
3D temperature at large radii due to an overly steep poor fit in the
deprojection procedure could lead to an even higher gas mass fraction
by driving the hydrostatic mass low.

For an abundance that is measured 5 times lower than the true
abundance, the measured gas mass fraction could be biased high by $70\%$
given a true abundance of ${Z}=0.05{Z}_{\odot}$.  

While the bias in enclosed gas mass and hydrostatic mass at small
radii is rather large, we do not consider this to be representative of
biases present in observations; the error in abundance measurement is
primarily an issue in the low density outskirts where photon
statistics are poor.  Additionally, at smaller radii, the gas is much
hotter, so the metal abundance and X-ray temperature parameters are
less degenerate with each other in the spectral fit.


\subsection{Potential issues for gas density slope measurements}
\label{sec:physical}


\begin{figure} [t]
  \centering
  \epsscale{1.1}
  \plotone{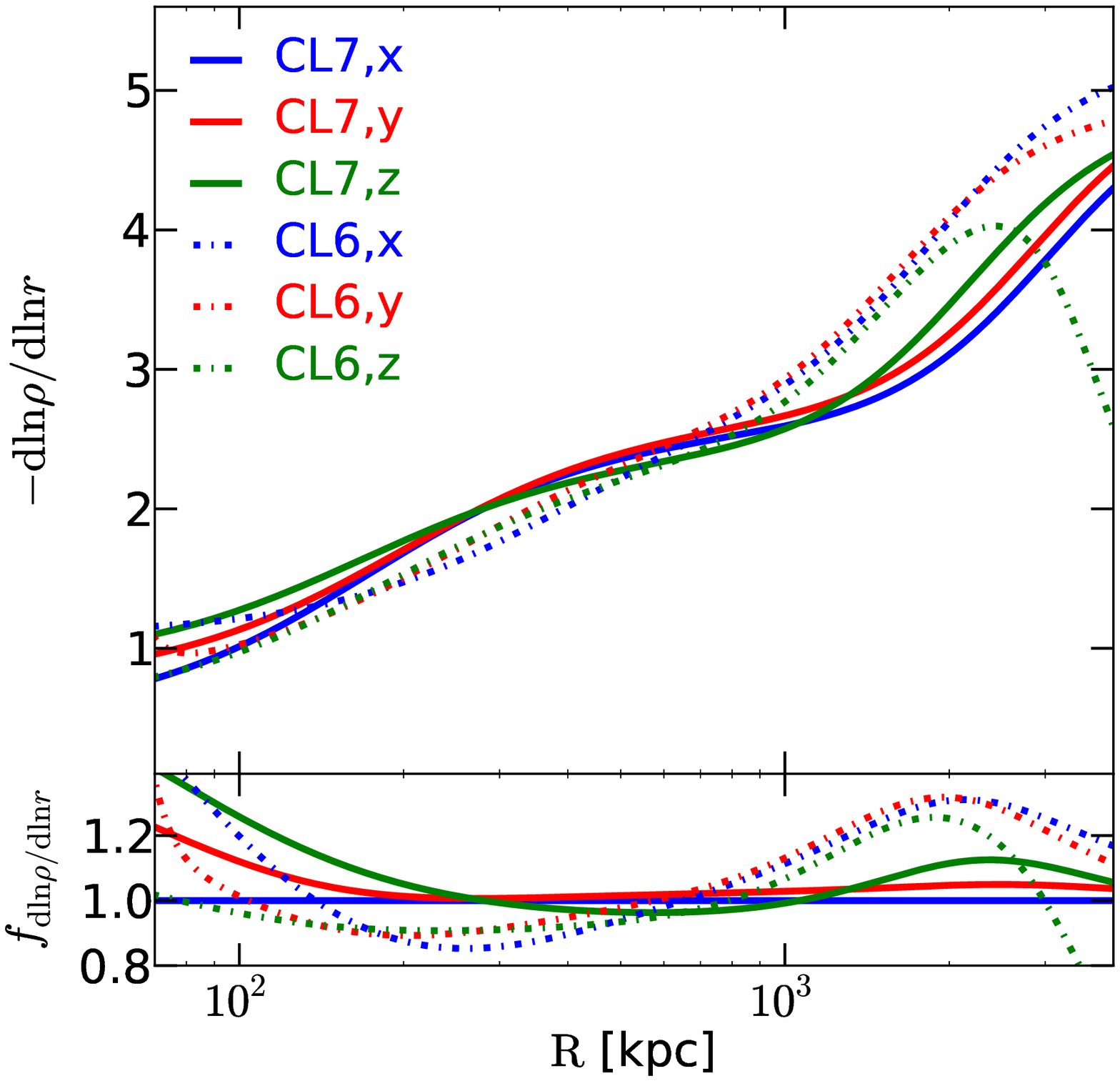}
  \caption{The logarithmic gas density slope profile of CL6
    (dash-dotted lines) and CL7 (solid lines) along different
    projections.  The z projection for CL6 is perpendicular to the
    plane of a merger. The bottom panel is the ratio of each profile
    to the profile of CL7 along the x-projection.  The logarithmic gas
    density slope of CL6, a dynamically disturbed cluster, is steeper
    than that of the CL7.}
  \label{fig:density_slope}
\end{figure}

We test two potential issues for density slope measurements in cluster
outskirts to compare with biases in measurements from spectral fitting
and metal abundance extrapolation: projection bias from triaxiality
and mass accretion history.

To test the bias due to projection effects from traxiality, we compute
the inertia tensors of the gas mass at $R_{500c}$ for our clusters.
For CL7, the semi-minor and semi-major axes of the inertia tensor lie
almost parallel to the $x$- and $y$-axes respectively, allowing us to
use the projections along these axes to make a preliminary assessment
of the effects of cluster shape. 

To test the effect of accretion history, we compare two clusters, CL6
and CL7 that have similar mass but differ in dynamical states.  CL6
has a recent major merger and CL7 has been quiescent since $z \sim 1$.
We compare these two clusters to identify the extent of steepening
that is mainly due to recent accretion history.  

Figure~\ref{fig:density_slope} shows the logarithmic gas density slope
for our two test cases along various projections, CL6 and CL7.  In our
relaxed cluster, CL7, the density slope projected along each axis
varies by no more than $15\%$. We would naively expect a significant
difference between a line of sight along the semi-major axis (y-axis)
and along the semi-minor axis (x-axis). However, the difference in
density slope between these two line of sight projections amounts to
only a few percent.  Effects due to line of sight projection are not
likely to significantly steepen observed density profiles.

CL6 is a dynamically disturbed cluster that experienced a 1:1 major
merger at $z\approx0.6$, and has a smaller infalling structure at
${R}\gsim R_{200c}$ at $z=0$.  The $x$- and $y$-projections correspond
to lines of sight that are perpendicular to the plane of accreting
substructure.  The effects of a merging substructure are visible in
the $z$-projection (dotted green), where the density slope is no
longer monotonically increasing at larger radii.  The outskirts of CL6
cluster along the $x$- and $y$-projections have an observed density
profile that exceeds the density slope along the $x$-projection of CL7
by $\approx30\%$, confirming that X-ray observations should be able to
see a difference in density slope between clusters of varying
accretion histories.

These two mechanisms have the potential to contribute additional bias
to the observed density slope than the bias introduced from spectral
fitting ($\approx5\%$). 


\section{Summary and Discussion}
\label{sec:summary}


The study of galaxy cluster outskirts is the new frontier in
cluster-based cosmology.  In this work, we used cosmological
simulations of galaxy clusters to investigate potential biases in
metal abundance measurements from X-ray spectra in the low density
cluster outskirts ($R>R_{500}$), where metal line cooling
contributions become significant.  Using X-ray maps with a constant
input metal abundance, we showed how systematically biased metal
abundance measurements affect ICM quantities such as the gas density
and temperature profiles.

In order to test other potential sources of systematic uncertainty in
density slope measurements from {\em  Chandra} observations of Abell 133
(Vikhlinin~et~al., in prep.), we performed our mock X-ray analysis for
projections along the semi-major and semi-minor axes of a relaxed
cluster, and for a dynamically disturbed cluster.

To summarize our findings: 

\begin{itemize}
\item Metal abundance measurements in the outskirts of our test case
  of a relaxed cluster shows that the metal abundance is best
  recovered to within a factor of two.  Due to the presence of a
  multi-temperature medium, strong metal line contributions will be
  systematically biased low.  In the regime of few photon counts
  (either from low X-ray surface brightness or low abundance),
  accurate metal line contributions will be difficult to recover from
  Poisson noise.
\item An incorrect measurement of the abundance in the low density
  regions corresponds to fluctuations in measured projected X-ray
  temperature to within $10\%$.  This temperature bias drives the
  projected emission measure in a direction opposite of the bias, thus
  affecting the gas density measurements as much as $15\%$ in the
  outskirts out to $\approx3\times R_{200c}$.
\item We test the extent of biases that may arise from incorrectly
  extrapolating the metal abundance measurement in galaxy cluster
  outskirts by fixing the metal abundance to a factor of 2 and a
  factor of 5 above and below the true abundance.  An abundance
  assumed a factor of 5 too high biases the gas density measurement
  and the enclosed total mass by as much as $15\%$ at ${R}_{200c}$.
\item In the cluster outskirts, an X-ray temperature that is biased
  low by $10\%$ corresponds to a decrease in the total mass
  measurement, increase in the enclosed gas mass, and therefore a
  boosted gas mass fraction by $\sim 10\%$.
\item Two potential physical mechanisms for the derived sharpening of
  density gradients in cluster outskirts are projection and dynamical
  state of the cluster.  Our test cases show that these contribute to
  no more than $15\%$ and $30\%$, respectively, to the logarithmic gas
  density slope. The bias due to an incorrect abundance measurement
  would affect the density slope by no more than $5\%$.
\end{itemize}

We note that fully testing the accuracy of X-ray measurements in
cluster outskirts requires a comparison of measurements of ICM
properties directly from the simulation with 3D filaments removed in a
self-consistent manner with our projected filament finder.  While the
test cases used in this study do not include high mass clusters, any
bias from incorrect abundance measurements would not exceed the bias
found in the lower mass counterparts.  High mass clusters have larger
temperatures at $R_{200c}$, resulting in a decreased relative
contribution of metal line emissions to the X-ray emission measure.
This analysis does not include any contributions from simulated
backgrounds, and the effects of instrumental and X-ray backgrounds are
additional sources of systematic biases that have not been accounted
for in our discussions.

Furthermore, our analysis does not take into account non-equilibrium
electrons at large radii, which have a lower temperature than the mean
ICM gas temperature \citep{ruddandnagai_09}.  The temperature bias
from non-equilibrium electrons is small in clusters of the mass scales
we tested, and we expect non-equilibrium electrons to have less of an
effect on the emission measure in the cluster outskirts than biases
from metal abundance measurements.  Metal line contributions have a
significant contribution to the X-ray emission in the low density
cluster outskirts, and the X-ray emission has a weak dependence on any
biases in the temperature at all radii ($\propto T^{1/2}$).
Alternatively, in higher mass clusters, non-equilibrium electrons have
larger deviations from the mean ICM gas temperature because the
equilibration timescale is longer.  The corresponding biases in X-ray
measurements of the outskirts of high mass clusters is a topic for
future work.

Our findings emphasize the need for long exposure times to have
accurate measurements of metal abundance and X-ray temperature in the
low density outskirts of galaxy clusters.  Accurate X-ray measurements
of the hydrostatic mass require an additional term in the
temperature modeling in cluster outskirts to account for the shallow
slope in the outermost radii. Additionally, accounting for errors in
the metal abundance measurement will require more careful modeling of
the thermal and chemical structure in cluster outskirts.

In addition to gas inhomogeneities \citep[e.g.,][]{nagaiandlau_11},
biased measurements of the metal abundance in cluster outskirts may
partially explain the enhanced gas mass fraction
\citep[e.g.,][]{simionescu_etal11} and flattened entropy profiles
\citep[e.g.,][]{walker_etal12} observed in cluster outskirts.  Errors
in metal abundance could produce enhanced emission measurements and a
steeper 3D temperature profile, leading to overestimates in gas mass
and underestimates in hydrostatic mass.

We have identified three potential mechanisms that contribute to
density steepening in our mock data, and may affect the observed
density slope beyond ${R}_{200c}$: (1) overestimation of the
abundance, (2) a line of sight corresponding to the semi-minor axis,
and (3) rapid recent mass accretion.  For (1), the biases introduced
by spectral fitting are not likely to exceed $5\%$.  The two physical
mechanisms (2) and (3) had respective effects of $15\%$ and $30\%$.

The mass accretion rate of a galaxy cluster can influence its DM
density \citep{cuesta_etal08,diemer_etal14} and gas profiles
(E.T.~Lau~et~al., in prep.).  Accurate measurements of gas density and
temperature profiles from X-ray observations of cluster outskirts will
provide information on the recent mass accretion history of a galaxy
cluster, which is one of the dominant systematics in mass-observable
relations in cluster cosmology.  Deep observations of cluster
outskirts will allow us to better understand the growth of galaxy
clusters through mass accretion from the surrounding filamentary
structures.

A careful consideration of all potential biases that enter into ICM
measurements in galaxy cluster outskirts will enable us to fully use
clusters as cosmological probes.  Results from this work will also be
useful for future missions, such as {\em SMART-X} \footnote{
  \url{http://smart-x.cfa.harvard.edu/} } and {\em Athena+}
\footnote{\url{http://athena2.irap.omp.eu/}}, which will begin mapping
the denser parts of the filamentary cosmic web.

\acknowledgments We thank the referee for the useful feedback, and
Elena Rasia for comments on the manuscript. This work is supported by
NSF grant AST-1009811, NASA ATP grant NNX11AE07G, NASA Chandra grants
GO213004B and TM4-15007X, and by the facilities and staff of the Yale
University Faculty of Arts and Sciences High Performance Computing
Center. CA acknowledges support from the NSF Graduate Student Research
Fellowship and the Alan D. Bromley Fellowship from Yale University.

\bibliography{ms}

\begin{thebibliography}{}
\expandafter\ifx\csname natexlab\endcsname\relax\def\natexlab#1{#1}\fi

\bibitem[{{Allen} {et~al.}(2008){Allen}, {Rapetti}, {Schmidt}, {Ebeling},
  {Morris}, \& {Fabian}}]{allen_etal08}
{Allen}, S.~W., {Rapetti}, D.~A., {Schmidt}, R.~W., {et~al.} 2008,
  \href{http://dx.doi.org/10.1111/j.1365-2966.2007.12610.x}{\mnras},
  \href{http://adsabs.harvard.edu/abs/2008MNRAS.383..879A}{383},
  \href{http://adsabs.harvard.edu/abs/2008MNRAS.383..879A}{879}

\bibitem[{{Anders} \& {Grevesse}(1989)}]{andersandgrevesse_89}
{Anders}, E., \& {Grevesse}, N. 1989,
  \href{http://dx.doi.org/10.1016/0016-7037(89)90286-X}{\gca},
  \href{http://adsabs.harvard.edu/abs/1989GeCoA..53..197A}{53},
  \href{http://adsabs.harvard.edu/abs/1989GeCoA..53..197A}{197}

\bibitem[{{Bautz} {et~al.}(2009){Bautz}, {Miller}, {Sanders}, {Arnaud},
  {Mushotzky}, {Porter}, {Hayashida}, {Henry}, {Hughes}, {Kawaharada},
  {Makashima}, {Sato}, \& {Tamura}}]{bautz_etal09}
{Bautz}, M.~W., {Miller}, E.~D., {Sanders}, J.~S., {et~al.} 2009, \pasj,
  \href{http://adsabs.harvard.edu/abs/2009PASJ...61.1117B}{61},
  \href{http://adsabs.harvard.edu/abs/2009PASJ...61.1117B}{1117}

\bibitem[{{B{\"o}hringer} \& {Werner}(2010)}]{bohringerandwerner_10}
{B{\"o}hringer}, H., \& {Werner}, N. 2010,
  \href{http://dx.doi.org/10.1007/s00159-009-0023-3}{\aapr},
  \href{http://adsabs.harvard.edu/abs/2010A%26ARv..18..127B}{18},
  \href{http://adsabs.harvard.edu/abs/2010A%26ARv..18..127B}{127}

\bibitem[{{Booth} \& {Schaye}(2009)}]{boothandschaye_09}
{Booth}, C.~M., \& {Schaye}, J. 2009,
  \href{http://dx.doi.org/10.1111/j.1365-2966.2009.15043.x}{\mnras},
  \href{http://adsabs.harvard.edu/abs/2009MNRAS.398...53B}{398},
  \href{http://adsabs.harvard.edu/abs/2009MNRAS.398...53B}{53}

\bibitem[{{Cash}(1979)}]{cash_79}
{Cash}, W. 1979, \href{http://dx.doi.org/10.1086/156922}{\apj},
  \href{http://adsabs.harvard.edu/abs/1979ApJ...228..939C}{228},
  \href{http://adsabs.harvard.edu/abs/1979ApJ...228..939C}{939}

\bibitem[{{Cuesta} {et~al.}(2008){Cuesta}, {Prada}, {Klypin}, \&
  {Moles}}]{cuesta_etal08}
{Cuesta}, A.~J., {Prada}, F., {Klypin}, A., \& {Moles}, M. 2008,
  \href{http://dx.doi.org/10.1111/j.1365-2966.2008.13590.x}{\mnras},
  \href{http://adsabs.harvard.edu/abs/2008MNRAS.389..385C}{389},
  \href{http://adsabs.harvard.edu/abs/2008MNRAS.389..385C}{385}

\bibitem[{{Diemer} \& {Kravtsov}(2014)}]{diemer_etal14}
{Diemer}, B., \& {Kravtsov}, A.~V. 2014, ArXiv e-prints,
  arXiv:\href{http://adsabs.harvard.edu/abs/2014arXiv1401.1216D}{1401.1216}

\bibitem[{{Eckert} {et~al.}(2013{\natexlab{a}}){Eckert}, {Ettori}, {Molendi},
  {Vazza}, \& {Paltani}}]{eckert_etal13a}
{Eckert}, D., {Ettori}, S., {Molendi}, S., {Vazza}, F., \& {Paltani}, S.
  2013{\natexlab{a}},
  \href{http://dx.doi.org/10.1051/0004-6361/201220403}{\aap},
  \href{http://adsabs.harvard.edu/abs/2013A%26A...551A..23E}{551},
  \href{http://adsabs.harvard.edu/abs/2013A%26A...551A..23E}{A23}

\bibitem[{{Eckert} {et~al.}(2013{\natexlab{b}}){Eckert}, {Molendi}, {Vazza},
  {Ettori}, \& {Paltani}}]{eckert_etal13b}
{Eckert}, D., {Molendi}, S., {Vazza}, F., {Ettori}, S., \& {Paltani}, S.
  2013{\natexlab{b}},
  \href{http://dx.doi.org/10.1051/0004-6361/201220402}{\aap},
  \href{http://adsabs.harvard.edu/abs/2013A%26A...551A..22E}{551},
  \href{http://adsabs.harvard.edu/abs/2013A%26A...551A..22E}{A22}

\bibitem[{{George} {et~al.}(2009){George}, {Fabian}, {Sanders}, {Young}, \&
  {Russell}}]{george_etal09}
{George}, M.~R., {Fabian}, A.~C., {Sanders}, J.~S., {Young}, A.~J., \&
  {Russell}, H.~R. 2009,
  \href{http://dx.doi.org/10.1111/j.1365-2966.2009.14547.x}{\mnras},
  \href{http://adsabs.harvard.edu/abs/2009MNRAS.395..657G}{395},
  \href{http://adsabs.harvard.edu/abs/2009MNRAS.395..657G}{657}

\bibitem[{{Hoshino} {et~al.}(2010){Hoshino}, {Henry}, {Sato}, {Akamatsu},
  {Yokota}, {Sasaki}, {Ishisaki}, {Ohashi}, {Bautz}, {Fukazawa}, {Kawano},
  {Furuzawa}, {Hayashida}, {Tawa}, {Hughes}, {Kokubun}, \&
  {Tamura}}]{hoshino_etal10}
{Hoshino}, A., {Henry}, J.~P., {Sato}, K., {et~al.} 2010, \pasj,
  \href{http://adsabs.harvard.edu/abs/2010PASJ...62..371H}{62},
  \href{http://adsabs.harvard.edu/abs/2010PASJ...62..371H}{371}

\bibitem[{{Kaastra} \& {Jansen}(1993)}]{kaastra_etal93}
{Kaastra}, J.~S., \& {Jansen}, F.~A. 1993, \aaps,
  \href{http://adsabs.harvard.edu/abs/1993A%26AS...97..873K}{97},
  \href{http://adsabs.harvard.edu/abs/1993A%26AS...97..873K}{873}

\bibitem[{{Kawaharada} {et~al.}(2010){Kawaharada}, {Okabe}, {Umetsu},
  {Takizawa}, {Matsushita}, {Fukazawa}, {Hamana}, {Miyazaki}, {Nakazawa}, \&
  {Ohashi}}]{kawaharada_etal10}
{Kawaharada}, M., {Okabe}, N., {Umetsu}, K., {et~al.} 2010,
  \href{http://dx.doi.org/10.1088/0004-637X/714/1/423}{\apj},
  \href{http://adsabs.harvard.edu/abs/2010ApJ...714..423K}{714},
  \href{http://adsabs.harvard.edu/abs/2010ApJ...714..423K}{423}

\bibitem[{{Klypin} {et~al.}(2001){Klypin}, {Kravtsov}, {Bullock}, \&
  {Primack}}]{klypin_etal01}
{Klypin}, A., {Kravtsov}, A.~V., {Bullock}, J.~S., \& {Primack}, J.~R. 2001,
  \href{http://dx.doi.org/10.1086/321400}{\apj},
  \href{http://adsabs.harvard.edu/abs/2001ApJ...554..903K}{554},
  \href{http://adsabs.harvard.edu/abs/2001ApJ...554..903K}{903}

\bibitem[{{Komiyama} {et~al.}(2009){Komiyama}, {Sato}, {Nagino}, {Ohashi}, \&
  {Matsushita}}]{komiyama_etal09}
{Komiyama}, M., {Sato}, K., {Nagino}, R., {Ohashi}, T., \& {Matsushita}, K.
  2009, \href{http://dx.doi.org/10.1093/pasj/61.sp1.S337}{\pasj},
  \href{http://adsabs.harvard.edu/abs/2009PASJ...61S.337K}{61},
  \href{http://adsabs.harvard.edu/abs/2009PASJ...61S.337K}{337}

\bibitem[{{Kravtsov}(1999)}]{kravtsov_99}
{Kravtsov}, A.~V. 1999, PhD thesis, New Mexico State Univ.

\bibitem[{{Kravtsov} {et~al.}(2002){Kravtsov}, {Klypin}, \&
  {Hoffman}}]{kravtsov_etal02}
{Kravtsov}, A.~V., {Klypin}, A., \& {Hoffman}, Y. 2002,
  \href{http://dx.doi.org/10.1086/340046}{\apj},
  \href{http://adsabs.harvard.edu/cgi-bin/nph-bib_query?bibcode=2002ApJ...571.%
.563K}{571},
  \href{http://adsabs.harvard.edu/cgi-bin/nph-bib_query?bibcode=2002ApJ...571.%
.563K}{563}

\bibitem[{{Leccardi} \& {Molendi}(2008)}]{leccardi_molendi08}
{Leccardi}, A., \& {Molendi}, S. 2008,
  \href{http://dx.doi.org/10.1051/0004-6361:200810113}{\aap},
  \href{http://adsabs.harvard.edu/abs/2008A%26A...487..461L}{487},
  \href{http://adsabs.harvard.edu/abs/2008A%26A...487..461L}{461}

\bibitem[{{Liedahl} {et~al.}(1995){Liedahl}, {Osterheld}, \&
  {Goldstein}}]{liedahl_etal95}
{Liedahl}, D.~A., {Osterheld}, A.~L., \& {Goldstein}, W.~H. 1995,
  \href{http://dx.doi.org/10.1086/187729}{\apjl},
  \href{http://adsabs.harvard.edu/abs/1995ApJ...438L.115L}{438},
  \href{http://adsabs.harvard.edu/abs/1995ApJ...438L.115L}{L115}

\bibitem[{{Matsushita} {et~al.}(2007){Matsushita}, {Fukazawa}, {Hughes},
  {Kitaguchi}, {Makishima}, {Nakazawa}, {Ohashi}, {Ota}, {Tamura}, {Tozuka},
  {Tsuru}, {Urata}, \& {Yamasaki}}]{matsushita_etal07}
{Matsushita}, K., {Fukazawa}, Y., {Hughes}, J.~P., {et~al.} 2007,
  \href{http://dx.doi.org/10.1093/pasj/59.sp1.S327}{\pasj},
  \href{http://adsabs.harvard.edu/abs/2007PASJ...59S.327M}{59},
  \href{http://adsabs.harvard.edu/abs/2007PASJ...59S.327M}{327}

\bibitem[{{Mazzotta} {et~al.}(2004){Mazzotta}, {Rasia}, {Moscardini}, \&
  {Tormen}}]{mazzotta_etal04}
{Mazzotta}, P., {Rasia}, E., {Moscardini}, L., \& {Tormen}, G. 2004,
  \href{http://dx.doi.org/10.1111/j.1365-2966.2004.08167.x}{\mnras},
  \href{http://adsabs.harvard.edu/abs/2004MNRAS.354...10M}{354},
  \href{http://adsabs.harvard.edu/abs/2004MNRAS.354...10M}{10}

\bibitem[{{Mewe} {et~al.}(1985){Mewe}, {Gronenschild}, \& {van den
  Oord}}]{mewe_etal85}
{Mewe}, R., {Gronenschild}, E.~H.~B.~M., \& {van den Oord}, G.~H.~J. 1985,
  \aaps, \href{http://adsabs.harvard.edu/abs/1985A%26AS...62..197M}{62},
  \href{http://adsabs.harvard.edu/abs/1985A%26AS...62..197M}{197}

\bibitem[{{Nagai} {et~al.}(2007{\natexlab{a}}){Nagai}, {Kravtsov}, \&
  {Vikhlinin}}]{nagai_etal07a}
{Nagai}, D., {Kravtsov}, A.~V., \& {Vikhlinin}, A. 2007{\natexlab{a}},
  \href{http://dx.doi.org/10.1086/521328}{\apj},
  \href{http://adsabs.harvard.edu/abs/2007ApJ...668....1N}{668},
  \href{http://adsabs.harvard.edu/abs/2007ApJ...668....1N}{1}

\bibitem[{{Nagai} \& {Lau}(2011)}]{nagaiandlau_11}
{Nagai}, D., \& {Lau}, E.~T. 2011,
  \href{http://dx.doi.org/10.1088/2041-8205/731/1/L10}{\apjl},
  \href{http://adsabs.harvard.edu/abs/2011ApJ...731L..10N}{731},
  \href{http://adsabs.harvard.edu/abs/2011ApJ...731L..10N}{L10}

\bibitem[{{Nagai} {et~al.}(2007{\natexlab{b}}){Nagai}, {Vikhlinin}, \&
  {Kravtsov}}]{nagai_etal07b}
{Nagai}, D., {Vikhlinin}, A., \& {Kravtsov}, A.~V. 2007{\natexlab{b}},
  \href{http://dx.doi.org/10.1086/509868}{\apj},
  \href{http://adsabs.harvard.edu/abs/2007ApJ...655...98N}{655},
  \href{http://adsabs.harvard.edu/abs/2007ApJ...655...98N}{98}

\bibitem[{{Rasia} {et~al.}(2008){Rasia}, {Mazzotta}, {Bourdin}, {Borgani},
  {Tornatore}, {Ettori}, {Dolag}, \& {Moscardini}}]{rasia_etal08}
{Rasia}, E., {Mazzotta}, P., {Bourdin}, H., {et~al.} 2008,
  \href{http://dx.doi.org/10.1086/524345}{\apj},
  \href{http://adsabs.harvard.edu/abs/2008ApJ...674..728R}{674},
  \href{http://adsabs.harvard.edu/abs/2008ApJ...674..728R}{728}

\bibitem[{{Reiprich} {et~al.}(2013){Reiprich}, {Basu}, {Ettori}, {Israel},
  {Lovisari}, {Molendi}, {Pointecouteau}, \& {Roncarelli}}]{reiprich_etal13}
{Reiprich}, T.~H., {Basu}, K., {Ettori}, S., {et~al.} 2013,
  \href{http://dx.doi.org/10.1007/s11214-013-9983-8}{\ssr},
  \href{http://adsabs.harvard.edu/abs/2013SSRv..177..195R}{177},
  \href{http://adsabs.harvard.edu/abs/2013SSRv..177..195R}{195}

\bibitem[{{Reiprich} {et~al.}(2009){Reiprich}, {Hudson}, {Zhang}, {Sato},
  {Ishisaki}, {Hoshino}, {Ohashi}, {Ota}, \& {Fujita}}]{reiprich_etal09}
{Reiprich}, T.~H., {Hudson}, D.~S., {Zhang}, Y.-Y., {et~al.} 2009,
  \href{http://dx.doi.org/10.1051/0004-6361/200810404}{\aap},
  \href{http://adsabs.harvard.edu/abs/2009A%26A...501..899R}{501},
  \href{http://adsabs.harvard.edu/abs/2009A%26A...501..899R}{899}

\bibitem[{{Roncarelli} {et~al.}(2013){Roncarelli}, {Ettori}, {Borgani},
  {Dolag}, {Fabjan}, \& {Moscardini}}]{roncarelli_etal13}
{Roncarelli}, M., {Ettori}, S., {Borgani}, S., {et~al.} 2013,
  \href{http://dx.doi.org/10.1093/mnras/stt654}{\mnras},
  \href{http://adsabs.harvard.edu/abs/2013MNRAS.432.3030R}{432},
  \href{http://adsabs.harvard.edu/abs/2013MNRAS.432.3030R}{3030}

\bibitem[{{Rudd} \& {Nagai}(2009)}]{ruddandnagai_09}
{Rudd}, D.~H., \& {Nagai}, D. 2009,
  \href{http://dx.doi.org/10.1088/0004-637X/701/1/L16}{\apjl},
  \href{http://adsabs.harvard.edu/abs/2009ApJ...701L..16R}{701},
  \href{http://adsabs.harvard.edu/abs/2009ApJ...701L..16R}{L16}

\bibitem[{{Rudd} {et~al.}(2008){Rudd}, {Zentner}, \& {Kravtsov}}]{rudd_etal08}
{Rudd}, D.~H., {Zentner}, A.~R., \& {Kravtsov}, A.~V. 2008,
  \href{http://dx.doi.org/10.1086/523836}{\apj},
  \href{http://adsabs.harvard.edu/abs/2008ApJ...672...19R}{672},
  \href{http://adsabs.harvard.edu/abs/2008ApJ...672...19R}{19}

\bibitem[{{Shakura} \& {Sunyaev}(1973)}]{shakuraandsunyaev_73}
{Shakura}, N.~I., \& {Sunyaev}, R.~A. 1973, \aap,
  \href{http://adsabs.harvard.edu/abs/1973A%26A....24..337S}{24},
  \href{http://adsabs.harvard.edu/abs/1973A%26A....24..337S}{337}

\bibitem[{{Simionescu} {et~al.}(2011){Simionescu}, {Allen}, {Mantz}, {Werner},
  {Takei}, {Morris}, {Fabian}, {Sanders}, {Nulsen}, {George}, \&
  {Taylor}}]{simionescu_etal11}
{Simionescu}, A., {Allen}, S.~W., {Mantz}, A., {et~al.} 2011,
  \href{http://dx.doi.org/10.1126/science.1200331}{Science},
  \href{http://adsabs.harvard.edu/abs/2011Sci...331.1576S}{331},
  \href{http://adsabs.harvard.edu/abs/2011Sci...331.1576S}{1576}

\bibitem[{{Vazza} {et~al.}(2013){Vazza}, {Eckert}, {Simionescu}, {Br{\"u}ggen},
  \& {Ettori}}]{vazza_etal13}
{Vazza}, F., {Eckert}, D., {Simionescu}, A., {Br{\"u}ggen}, M., \& {Ettori}, S.
  2013, \href{http://dx.doi.org/10.1093/mnras/sts375}{\mnras},
  \href{http://adsabs.harvard.edu/abs/2013MNRAS.429..799V}{429},
  \href{http://adsabs.harvard.edu/abs/2013MNRAS.429..799V}{799}

\bibitem[{{Vikhlinin} {et~al.}(2006){Vikhlinin}, {Kravtsov}, {Forman}, {Jones},
  {Markevitch}, {Murray}, \& {Van Speybroeck}}]{vikhlinin_etal06}
{Vikhlinin}, A., {Kravtsov}, A., {Forman}, W., {et~al.} 2006,
  \href{http://dx.doi.org/10.1086/500288}{\apj},
  \href{http://adsabs.harvard.edu/abs/2006ApJ...640..691V}{640},
  \href{http://adsabs.harvard.edu/abs/2006ApJ...640..691V}{691}

\bibitem[{{Vikhlinin} {et~al.}(2005){Vikhlinin}, {Markevitch}, {Murray},
  {Jones}, {Forman}, \& {Van Speybroeck}}]{vikhlinin_etal05}
{Vikhlinin}, A., {Markevitch}, M., {Murray}, S.~S., {et~al.} 2005,
  \href{http://dx.doi.org/10.1086/431142}{\apj},
  \href{http://adsabs.harvard.edu/abs/2005ApJ...628..655V}{628},
  \href{http://adsabs.harvard.edu/abs/2005ApJ...628..655V}{655}

\bibitem[{{Vikhlinin} {et~al.}(1998){Vikhlinin}, {McNamara}, {Forman}, {Jones},
  {Quintana}, \& {Hornstrup}}]{vikhlinin_etal98}
{Vikhlinin}, A., {McNamara}, B.~R., {Forman}, W., {et~al.} 1998,
  \href{http://dx.doi.org/10.1086/305951}{\apj},
  \href{http://adsabs.harvard.edu/abs/1998ApJ...502..558V}{502},
  \href{http://adsabs.harvard.edu/abs/1998ApJ...502..558V}{558}

\bibitem[{{Vikhlinin} {et~al.}(2009){Vikhlinin}, {Kravtsov}, {Burenin},
  {Ebeling}, {Forman}, {Hornstrup}, {Jones}, {Murray}, {Nagai}, {Quintana}, \&
  {Voevodkin}}]{vikhlinin_etal09}
{Vikhlinin}, A., {Kravtsov}, A.~V., {Burenin}, R.~A., {et~al.} 2009,
  \href{http://dx.doi.org/10.1088/0004-637X/692/2/1060}{\apj},
  \href{http://adsabs.harvard.edu/abs/2009ApJ...692.1060V}{692},
  \href{http://adsabs.harvard.edu/abs/2009ApJ...692.1060V}{1060}

\bibitem[{{Walker} {et~al.}(2012){Walker}, {Fabian}, {Sanders}, \&
  {George}}]{walker_etal12}
{Walker}, S.~A., {Fabian}, A.~C., {Sanders}, J.~S., \& {George}, M.~R. 2012,
  \href{http://dx.doi.org/10.1111/j.1745-3933.2012.01342.x}{\mnras},
  \href{http://adsabs.harvard.edu/abs/2012MNRAS.427L..45W}{427},
  \href{http://adsabs.harvard.edu/abs/2012MNRAS.427L..45W}{L45}

\bibitem[{{Werner} {et~al.}(2013){Werner}, {Urban}, {Simionescu}, \&
  {Allen}}]{werner_etal13}
{Werner}, N., {Urban}, O., {Simionescu}, A., \& {Allen}, S.~W. 2013,
  \href{http://dx.doi.org/10.1038/nature12646}{\nat},
  \href{http://adsabs.harvard.edu/abs/2013Natur.502..656W}{502},
  \href{http://adsabs.harvard.edu/abs/2013Natur.502..656W}{656}

\bibitem[{{Zhuravleva} {et~al.}(2013){Zhuravleva}, {Churazov}, {Kravtsov},
  {Lau}, {Nagai}, \& {Sunyaev}}]{zhuravleva_etal13}
{Zhuravleva}, I., {Churazov}, E., {Kravtsov}, A., {et~al.} 2013,
  \href{http://dx.doi.org/10.1093/mnras/sts275}{\mnras},
  \href{http://adsabs.harvard.edu/abs/2013MNRAS.428.3274Z}{428},
  \href{http://adsabs.harvard.edu/abs/2013MNRAS.428.3274Z}{3274}

\end{thebibliography}

\end{document}